\documentclass[11pt,a4paper]{article}
\usepackage[T1]{fontenc}
\usepackage{setspace}
\usepackage[latin1]{inputenc}
\usepackage{amsmath}
\usepackage{amssymb}
\usepackage{setspace}
\usepackage{esint}

\date{}

\allowdisplaybreaks

\numberwithin{equation}{section}

\title{Nilpotency of the $b$ ghost in the non-minimal pure spinor formalism}

\author{Renann Lipinski Jusinskas\thanks{renannlj@ift.unesp.br%
}}

\begin{document}

\maketitle
\begin{center}
ICTP South American Institute for Fundamental Research \\
Instituto de F\'isica Te\'orica, UNESP - Univ. Estadual Paulista \\
Rua Dr. Bento T. Ferraz 271, 01140-070, S\~ao Paulo, SP, Brasil.
\par\end{center}

\

\begin{abstract}The $b$ ghost in the non-minimal pure spinor formalism is not a fundamental
field. It is based on a complicated chain of operators and proving
its nilpotency is nontrivial. Chandia proved this property in arXiv:1008.1778,
but with an assumption on the non-minimal variables that is not valid
in general. In this work, the $b$ ghost is demonstrated to be nilpotent
without this assumption.\end{abstract}

\section{Introduction}

\

The super Poincar\'e covariant quantization of the superstring was achieved
in the year 2000, with the development of the pure spinor formalism
\cite{Berkovits:2000fe}. One of its oddest features is the absence
of a natural prescription for string loop amplitudes, that is manifest
in the other formalisms of bosonic and supersymmetric strings, due
to the existence of the world-sheet reparametrization invariance.

It is a well known fact that in gauge fixing the reparametrization
symmetry, a $\left(b,c\right)$ system rises as the ghost-antighost
pair. The $c$ ghost is a conformal weight $-1$ field, as it comes
from the general coordinate transformation parameter, and the $b$
ghost, the conjugate of $c$, is a conformal weight $+2$ field.

Concerning amplitudes, the fundamental objects of study in quantum
strings, the $c$ ghost appears at tree and $1$-loop level. In these
world-sheet topologies (respectively, the sphere and the torus), the
conformal Killing symmetries can be removed by fixing some vertex
positions. For the pure spinor formalism, Berkovits developed a prescription
\cite{Berkovits:2000fe,Berkovits:2004px,Berkovits:2005bt} that successfully
described superstring amplitudes, where a possible $c$ ghost played
no role at all.

For the $b$ ghost, however, the story is different. In a BRST-like
description, $b$ ghost insertions lie in the heart of the BRST invariance
of string loop amplitudes. The fundamental property is $\left\{ Q,b\right\} =T$,
where $T$ is the energy-momentum tensor (since the BRST charge has
ghost number $+1$, the $b$ ghost must have ghost number $-1$).
Combined with the Beltrami differentials, this property induces only
a surface contribution in the \emph{moduli} space integration.

In the minimal pure spinor formalism, where the available ghost variables
are the pure spinor $\lambda^{\alpha}$ and its conjugate $\omega_{\alpha}$,
the $b$ ghost is based upon a complicated chain of operators and
can be implemented only in a picture raised manner \cite{Berkovits:2004px},
as there are no suitable ghost number $-1$ fields.

With the addition of the ghost fields $\left(\overline{\lambda}_{\alpha},r_{\alpha}\right)$
and their conjugates $\left(\overline{\omega}^{\alpha},s^{\alpha}\right)$,
the non-minimal pure spinor formalism enables a much simpler construction
of the $b$ ghost \cite{Berkovits:2005bt}. More than that, the theory
can be interpreted as a twisted $\mathcal{N}=2$ $\hat{c}=3$ topological
string, where the BRST charge and the $b$ ghost are the fermionic
generators, while the ghost number current and the energy-momentum
tensor are the bosonic ones. This fact allowed the covariant computation
of multiloop superstring amplitudes without picture changing operators,
making the super Poincar\'e symmetry explicit in all the steps.

Since the $b$ ghost is a composite field, its nilpotency, a crucial
property in the topological string interpretation, is not evident.
In \cite{Chandia:2010ix}, the regularity of the $b$ ghost OPE with
itself was derived, but in an incomplete manner%
\footnote{The flaw in the proof of \cite{Chandia:2010ix} was pointed out by
N. Berkovits, in a private communication.%
}, as will be explained here. Therefore, a rigorous proof of such a
fundamental property was still lacking.

This paper is organized as follows. Section \ref{sec:Review} contains
a review of the pure spinor formalisms and section \ref{sec:bghost}
presents the construction of the $b$ ghost, its basic properties
and a rigorous derivation of the $b$ ghost OPE with itself, proving
its regularity. Appendix \ref{sec:Conventions} contains the conventions
that are being used in this work and some ordering considerations.

\section{Review of the pure spinor formalism\label{sec:Review}}

\

The pure spinor formalism will be reviewed here, establishing the
fundamental fields that will be used in the remaining sections.

\subsection{Matter fields}

\

The matter part of the action is
\begin{equation}
S_{m}=\frac{1}{2\pi}\int d^{2}z\left(\frac{1}{\alpha'}\partial X^{m}\overline{\partial}X_{m}+p_{\beta}\overline{\partial}\theta^{\beta}\right),
\end{equation}
and the free field propagators are just\begin{subequations}
\begin{eqnarray}
X^{m}\left(z,\bar{z}\right)X^{n}\left(y,\bar{y}\right) & \sim & -\frac{\alpha'}{2}\eta^{mn}\ln\left|z-y\right|^{2}\\
p_{\alpha}\left(z\right)\theta^{\beta}\left(y\right) & \sim & \frac{\delta_{\alpha}^{\beta}}{z-y}.
\end{eqnarray}
\end{subequations}

The action $S_{m}$ is invariant under the supersymmetric charge,
defined as
\begin{equation}
q_{\alpha}=\oint\left[p_{\alpha}+\frac{1}{\alpha'}\partial X^{m}\left(\theta\gamma_{m}\right)_{\alpha}+\frac{1}{12\alpha'}\left(\theta\gamma_{m}\partial\theta\right)\left(\theta\gamma_{m}\right)_{\alpha}\right].
\end{equation}
Note that
\begin{equation}
\alpha'\left\{ q_{\alpha},q_{\beta}\right\} =2\gamma_{\alpha\beta}^{m}\oint\partial X_{m}.
\end{equation}
The construction of the supersymmetric invariants follows:\begin{subequations}
\begin{eqnarray}
\Pi^{m} & = & \partial X^{m}+\frac{1}{2}\left(\theta\gamma^{m}\partial\theta\right),\\
d_{\alpha} & = & p_{\alpha}-\frac{1}{\alpha'}\partial X^{m}\left(\theta\gamma_{m}\right)_{\alpha}-\frac{1}{4\alpha'}\left(\theta\gamma^{m}\partial\theta\right)\left(\theta\gamma_{m}\right)_{\alpha}.
\end{eqnarray}
\end{subequations}So far, this is nothing but the left-moving sector
of the Green-Schwarz-Siegel action in the conformal gauge. The Virasoro
constraint is $\Pi^{m}\Pi_{m}+\alpha'd_{\alpha}\partial\theta^{\alpha}=0$
and the fermionic constraints (related to \emph{kappa} symmetry) are
$d_{\alpha}=0$.

The related OPE's are given by:\begin{subequations}
\begin{eqnarray}
\Pi^{m}\left(z\right)\Pi^{n}\left(y\right) & \sim & -\frac{\alpha'}{2}\frac{\eta^{mn}}{\left(z-y\right)^{2}},\\
d_{\alpha}\left(z\right)\Pi^{m}\left(y\right) & \sim & \frac{\gamma_{\alpha\beta}^{m}\partial\theta^{\beta}}{\left(z-y\right)},\\
d_{\alpha}\left(z\right)d_{\beta}\left(y\right) & \sim & -\frac{2}{\alpha'}\frac{\gamma_{\alpha\beta}^{m}\Pi_{m}}{\left(z-y\right)}.
\end{eqnarray}
\end{subequations}

The matter energy-momentum tensor (Virasoro constraint) is
\begin{equation}
T_{\textrm{matter}}=-\frac{1}{\alpha'}\partial X^{m}\partial X_{m}-p_{\alpha}\partial\theta^{\alpha},
\end{equation}
from which it follows that
\begin{equation}
T_{\textrm{matter}}\left(z\right)T_{\textrm{matter}}\left(y\right)\sim-\frac{11}{\left(z-y\right)^{4}}+2\frac{T_{\textrm{matter}}}{\left(z-y\right)^{2}}+\frac{\partial T_{\textrm{matter}}}{\left(z-y\right)}.\label{eq:opeTTmatter}
\end{equation}
Therefore, the free matter action yields a negative central charge,
that will be cancelled with the contribution coming from the ghost
sector.

\subsection{Ghost fields\label{sub:Ghost-fields}}

\

Introducing a pure spinor $\lambda^{\alpha}$ variable and its conjugate,
$\omega_{\alpha}$, one is able to define
\begin{equation}
J_{\textrm{BRST}}\equiv\lambda^{\alpha}d_{\alpha},\label{eq:brstmin}
\end{equation}
and construct a BRST like charge,
\begin{equation}
Q=\oint J_{\textrm{BRST}},\label{eq:brstchargemin}
\end{equation}
where
\begin{equation}
\left\{ Q,Q\right\} =-\frac{2}{\alpha'}\oint\left(\lambda\gamma^{m}\lambda\right)\Pi_{m}=0,
\end{equation}
and
\begin{equation}
\lambda\gamma^{m}\lambda=0\label{eq:psconstraint}
\end{equation}
 is the $D=10$ pure spinor constraint, which implies that only $11$
components of $\lambda^{\alpha}$ are independent. Observe that an
explicitly Lorentz invariant action for the ghost sector,
\begin{equation}
S_{\lambda}=\frac{1}{2\pi}\int d^{2}z\left(\omega_{\alpha}\bar{\partial}\lambda^{\alpha}\right),\label{eq:actionminps}
\end{equation}
must be gauge invariant under $\delta_{\epsilon}\omega_{\alpha}=\epsilon_{m}\left(\gamma^{m}\lambda\right)_{\alpha}$,
due to \eqref{eq:psconstraint}.

The simplest gauge invariant objects are
\begin{eqnarray*}
T_{\lambda}=-\omega\partial\lambda, & N^{mn}=-\frac{1}{2}\omega\gamma^{mn}\lambda, & J_{\lambda}=-\omega\lambda,
\end{eqnarray*}
respectively, the energy-momentum tensor, the Lorentz current and
the ghost number current.

The full set of OPE's of the ghost sector is:
\[
\begin{array}{ccc}
T_{\lambda}\left(z\right)T_{\lambda}\left(y\right)\sim\frac{11}{\left(z-y\right)^{4}}+2\frac{T_{\lambda}}{\left(z-y\right)^{2}}+\frac{\partial T_{\lambda}}{\left(z-y\right)}, &  & J_{\lambda}\left(z\right)T_{\lambda}\left(y\right)\sim-\frac{8}{\left(z-y\right)^{3}}+\frac{J_{\lambda}}{\left(z-y\right)^{2}},\end{array}
\]
\[
\begin{array}{ccccc}
N^{mn}\left(z\right)T_{\lambda}\left(y\right)\sim\frac{N^{mn}}{\left(z-y\right)^{2}}, &  & T_{\lambda}\left(z\right)\lambda^{\alpha}\left(y\right)\sim\frac{\partial\lambda^{\alpha}}{\left(z-y\right)}, &  & N^{mn}\left(z\right)\lambda^{\alpha}\left(y\right)\sim\frac{1}{2}\frac{\left(\gamma^{mn}\lambda\right)^{\alpha}}{\left(z-y\right)},\end{array}
\]
\[
\begin{array}{ccccc}
N^{mn}\left(z\right)J_{\lambda}\left(y\right)\sim\textrm{regular}, &  & J_{\lambda}\left(z\right)\lambda^{\alpha}\left(y\right)\sim\frac{\lambda^{\alpha}}{\left(z-y\right)}, &  & J_{\lambda}\left(z\right)J_{\lambda}\left(y\right)\sim-\frac{4}{\left(z-y\right)^{2}},\end{array}
\]
\[
N^{mn}\left(z\right)N^{pq}\left(y\right)\sim6\frac{\eta^{m[p}\eta^{q]n}}{\left(z-y\right)^{2}}+2\frac{\eta^{m[q}N^{p]n}+\eta^{n[p}N^{q]m}}{\left(z-y\right)}.
\]

The non-minimal version of the pure spinor formalism includes a new
set of fields, $\left(\overline{\lambda}_{\alpha},r_{\alpha}\right)$.
The former is also a pure spinor, that is
\begin{equation}
\overline{\lambda}\gamma^{m}\overline{\lambda}=0,\label{eq:nonpsconstraint}
\end{equation}
whereas the latter is a fermionic spinor constrained through
\begin{equation}
\overline{\lambda}\gamma^{m}r=0.\label{eq:lambdabarrconstraint}
\end{equation}
Both constraints imply that there are only $11$ independent components
in each spinor. Denoting their conjugates as $\left(\overline{\omega}^{\alpha},s^{\alpha}\right)$,
the action for the non-minimal sector is
\begin{equation}
S_{\overline{\lambda}}=\frac{1}{2\pi}\int d^{2}z\left(\bar{\omega}^{\alpha}\bar{\partial}\bar{\lambda}_{\alpha}+s^{\alpha}\bar{\partial}r_{\alpha}\right),\label{eq:actionnmps}
\end{equation}
which is gauge invariant by the following transformations
\begin{eqnarray}
\delta_{\epsilon,\phi}\bar{\omega}^{\alpha} & = & \epsilon^{m}\left(\gamma_{m}\bar{\lambda}\right)^{\alpha}+\phi^{m}\left(\gamma_{m}r\right)^{\alpha},\nonumber \\
\delta_{\phi}s^{\alpha} & = & \phi^{m}\left(\gamma_{m}\bar{\lambda}\right)^{\alpha}.\label{eq:nonminimalgaugeinv}
\end{eqnarray}

There are several gauge invariant quantities that can be built out
of $\overline{\omega}^{\alpha}$ and $s^{\alpha}$.

\begin{equation}
\begin{array}{ccc}
 & \overline{N}^{mn}=\frac{1}{2}\left(\bar{\lambda}\gamma^{mn}\bar{\omega}-r\gamma^{mn}s\right),\\
J_{\bar{\lambda}}=-\bar{\lambda}\bar{\omega}, & T_{\bar{\lambda}}=-\bar{\omega}\partial\bar{\lambda}-s\partial r, & \Phi=r\bar{\omega},\\
S=\overline{\lambda}s, & S^{mn}=\frac{1}{2}\bar{\lambda}\gamma^{mn}s, & J_{r}=rs.
\end{array}\label{eq:nonminimalcurrents}
\end{equation}
Here, $\overline{N}^{mn}$ is the Lorentz generator, $T_{\overline{\lambda}}$
is the energy-momentum tensor, and $J_{\bar{\lambda}}$ and $J_{r}$
are the ghost number currents. Note that they are not all independent\footnote{Some numerical coefficients of the corresponding relations in \cite{Berkovits:2005bt}
are incorrect, as can be promptly verified by the definitions in \eqref{eq:nonminimalcurrents}
and the gamma matrices identity \eqref{eq:gammaid1}.%
},
since
\begin{equation}
S^{mn}\left(\frac{r\gamma_{mn}\lambda}{\overline{\lambda}\lambda}\right)+S\left(\frac{r\lambda}{\overline{\lambda}\lambda}\right)-4J_{r}=0,
\end{equation}
\begin{equation}
\overline{N}^{mn}\left(\frac{r\gamma_{mn}\lambda}{\overline{\lambda}\lambda}\right)-J_{\overline{\lambda}}\left(\frac{r\lambda}{\overline{\lambda}\lambda}\right)+3J_{r}\left(\frac{r\lambda}{\overline{\lambda}\lambda}\right)+4\Phi=0.
\end{equation}

The OPE's between them can be summarized as follows:
\[
\begin{array}{ccc}
T_{\overline{\lambda}}\left(z\right)T_{\overline{\lambda}}\left(y\right)\sim2\frac{T_{\overline{\lambda}}}{\left(z-y\right)^{2}}+\frac{\partial T_{\overline{\lambda}}}{\left(z-y\right)}, & \overline{N}^{mn}\left(z\right)T_{\overline{\lambda}}\left(y\right)\sim\frac{\overline{N}^{mn}}{\left(z-y\right)^{2}}, & S^{mn}\left(z\right)T_{\overline{\lambda}}\left(y\right)\sim\frac{S^{mn}}{\left(z-y\right)^{2}},\end{array}
\]
\[
\begin{array}{ccc}
J_{\bar{\lambda}}\left(z\right)T_{\overline{\lambda}}\left(y\right)\sim-\frac{11}{\left(z-y\right)^{3}}+\frac{\overline{J}_{\overline{\lambda}}}{\left(z-y\right)^{2}}, & \Phi\left(z\right)T_{\overline{\lambda}}\left(y\right)\sim\frac{\Phi}{\left(z-y\right)^{2}}, & S\left(z\right)T_{\overline{\lambda}}\left(y\right)\sim\frac{S}{\left(z-y\right)^{2}},\end{array}
\]
\[
\begin{array}{ccc}
J_{r}\left(z\right)T_{\overline{\lambda}}\left(y\right)\sim\frac{11}{\left(z-y\right)^{3}}+\frac{J_{r}}{\left(z-y\right)^{2}}, & T_{\overline{\lambda}}\left(z\right)\overline{\lambda}_{\alpha}\left(y\right)\sim\frac{\partial\overline{\lambda}_{\alpha}}{\left(z-y\right)}, & T_{\overline{\lambda}}\left(z\right)r_{\alpha}\left(y\right)\sim\frac{\partial r_{\alpha}}{\left(z-y\right)},\end{array}
\]
\[
\begin{array}{ccc}
\Phi\left(z\right)S\left(y\right)\sim-\frac{8}{\left(z-y\right)^{2}}-\frac{J_{\overline{\lambda}}+J_{r}}{\left(z-y\right)}, & \Phi\left(z\right)S^{mn}\left(y\right)\sim\frac{\overline{N}^{mn}}{\left(z-y\right)}, & \Phi\left(z\right)\overline{\lambda}_{\alpha}\left(y\right)\sim-\frac{r_{\alpha}}{\left(z-y\right)},\end{array}
\]
\[
\begin{array}{ccc}
\Phi\left(z\right)\Phi\left(y\right)\sim\textrm{regular}, & \overline{N}^{mn}\left(z\right)J_{\bar{\lambda}}\left(y\right)\sim\textrm{regular}, & \overline{N}^{mn}\left(z\right)\Phi\left(y\right)\sim\textrm{regular},\end{array}
\]
\[
\begin{array}{ccc}
\overline{N}^{mn}\left(z\right)\overline{N}^{pq}\left(y\right)\sim2\frac{\eta^{m[q}\overline{N}^{p]n}+\eta^{n[p}\overline{N}^{q]m}}{\left(z-y\right)}, & J_{\bar{\lambda}}\left(z\right)J_{r}\left(y\right)\sim-\frac{3}{\left(z-y\right)^{2}}, & J_{\bar{\lambda}}\left(z\right)J_{\bar{\lambda}}\left(y\right)\sim-\frac{5}{\left(z-y\right)^{2}},\end{array}
\]
\[
\begin{array}{ccc}
\overline{N}^{mn}\left(z\right)J_{r}\left(y\right)\sim\textrm{regular}, & \overline{N}^{mn}\left(z\right)S\left(y\right)\sim\textrm{regular}, & J_{r}\left(z\right)J_{r}\left(y\right)\sim\frac{11}{\left(z-y\right)^{2}},\end{array}
\]
\[
\begin{array}{ccc}
\overline{N}^{mn}\left(z\right)\overline{\lambda}_{\alpha}\left(y\right)\sim-\frac{1}{2}\frac{\left(\overline{\lambda}\gamma^{mn}\right)_{\alpha}}{\left(z-y\right)}, & \overline{N}^{mn}\left(z\right)r_{\alpha}\left(y\right)\sim-\frac{1}{2}\frac{\left(r\gamma^{mn}\right)_{\alpha}}{\left(z-y\right)}, & J_{\bar{\lambda}}\left(z\right)\overline{\lambda}_{\alpha}\left(y\right)\sim\frac{\overline{\lambda}_{\alpha}}{\left(z-y\right)},\end{array}
\]
\[
\begin{array}{ccc}
J_{r}\left(z\right)r_{\alpha}\left(y\right)\sim\frac{r_{\alpha}\left(y\right)}{\left(z-y\right)}, & J_{\bar{\lambda}}\left(z\right)r_{\alpha}\left(y\right)\sim\textrm{regular}, & J_{r}\left(z\right)\overline{\lambda}_{\alpha}\left(y\right)\sim\textrm{regular}.\end{array}
\]
Note that there are no contributions to the central charge or to the
level of the Lorentz algebra%
\footnote{The quadratic pole in $\Phi\left(z\right)S\left(y\right)$ is also
absent in \cite{Berkovits:2005bt}. %
}.

The non-minimal variables enter the formalism in a very simple way, as
the BRST charge is defined to be
\begin{equation}
Q\equiv\oint\underbrace{\left(\lambda^{\alpha}d_{\alpha}+\Phi\right)}_{J_{BRST}\left(z\right)}.\label{eq:BRSTnonminimal}
\end{equation}
The same notation was used for the BRST charge in the minimal formalism,
but from now on, only \eqref{eq:BRSTnonminimal} will be referred
to as $Q$. The cohomology of \eqref{eq:BRSTnonminimal} is independent
of $\left(\overline{\lambda},\overline{\omega},r,s\right)$, as can
be seen from the quartet argument, and there is a state $\xi$ that
trivializes it:
\begin{equation}
\begin{array}{ccc}
\xi=\frac{\overline{\lambda}\cdot\theta}{\overline{\lambda}\cdot\lambda-r\cdot\theta}, &  & \left\{ Q,\xi\right\} =1.\end{array}
\end{equation}
Since $r_{\alpha}$ and $\theta^{\alpha}$ are grassmannian variables,
$\xi$ can be expanded as a finite power series in terms of $r\cdot\theta$.
Besides, $r_{\alpha}$ has only $11$ independent components, in such
a way that
\begin{equation}
\xi=\frac{\overline{\lambda}\cdot\theta}{\overline{\lambda}\cdot\lambda}\sum_{n=0}^{11}\left(\frac{r\cdot\theta}{\overline{\lambda}\cdot\lambda}\right)^{n}.
\end{equation}
Therefore, one way of avoiding the appearance of $\xi$ is limiting
the amount of inverse powers of $\overline{\lambda}\lambda$. However,
this is a fundamental ingredient in the construction of the $b$ ghost,
constituting the main obstruction for loop amplitude calculations
in the pure spinor formalism \cite{Berkovits:2005bt,Berkovits:2006vi}.

\section{The $b$ ghost\label{sec:bghost}}

\

The $b$ ghost is a central field in string perturbation theory, being
related to the BRST invariance of string loop amplitudes. Its basic
property is
\begin{equation}
\left\{ Q,b\right\} =T.
\end{equation}
As there is not such a fundamental object in the pure spinor formalism,
it must be build out of the available fields of the theory.

As introduced in \cite{Berkovits:2005bt}, the construction of the
non-minimal $b$ ghost is based on a chain of operators satisfying
some special relations, that will be reviewed below.

\subsection{Definition and properties}

\

The\emph{ }full quantum version\emph{ }of the $b$ ghost can be cast
as
\begin{equation}
b=b_{-1}+b_{0}+b_{1}+b_{2}+b_{3},\label{eq:quantumb}
\end{equation}
where\begin{subequations}
\begin{eqnarray}
b_{-1} & \equiv & -s^{\alpha}\partial\overline{\lambda}_{\alpha},\\
b_{0} & \equiv & \left(\frac{\overline{\lambda}_{\alpha}}{\left(\overline{\lambda}\lambda\right)},G^{\alpha}\right)+O,\\
b_{1} & \equiv & -2!\left(\frac{\overline{\lambda}_{\alpha}r_{\beta}}{\left(\overline{\lambda}\lambda\right)^{2}},H^{\alpha\beta}\right),\\
b_{2} & \equiv & -3!\left(\frac{\overline{\lambda}_{\alpha}r_{\beta}r_{\gamma}}{\left(\overline{\lambda}\lambda\right)^{3}},K^{\alpha\beta\gamma}\right),\\
b_{3} & \equiv & 4!\left(\frac{\overline{\lambda}_{\alpha}r_{\beta}r_{\gamma}r_{\lambda}}{\left(\overline{\lambda}\lambda\right)^{4}},L^{\alpha\beta\gamma\lambda}\right),
\end{eqnarray}
\end{subequations}and\begin{subequations}
\begin{eqnarray}
O & \equiv & -\partial\left(\frac{\overline{\lambda}_{\alpha}\overline{\lambda}_{\beta}}{\left(\overline{\lambda}\lambda\right)^{2}}\right)\lambda^{\alpha}\partial\theta^{\beta},\\
G^{\alpha} & = & \frac{1}{2}\gamma_{m}^{\alpha\beta}\left(\Pi^{m},d_{\beta}\right)-\frac{1}{4}N_{mn}\left(\gamma^{mn}\partial\theta\right)^{\alpha}-\frac{1}{4}J_{\lambda}\partial\theta^{\alpha}+4\partial^{2}\theta^{\alpha},\\
H^{\alpha\beta} & = & \frac{1}{4\cdot96}\gamma_{mnp}^{\alpha\beta}\left(\frac{\alpha'}{2}d\gamma^{mnp}d+24N^{mn}\Pi^{p}\right),\\
K^{\alpha\beta\gamma} & = & -\frac{1}{96}\left(\frac{\alpha'}{2}\right)N_{mn}\gamma_{mnp}^{[\alpha\beta}\left(\gamma^{p}d\right)^{\gamma]},\\
L^{\alpha\beta\gamma\lambda} & = & -\frac{3}{\left(96\right)^{2}}\left(\frac{\alpha'}{2}\right)\left(N^{mn},N^{rs}\right)\eta^{pq}\gamma_{mnp}^{[\alpha\beta}\gamma_{qrs}^{\gamma]\lambda}.
\end{eqnarray}
\end{subequations}Note that the subscript $n$ in $b_{n}$ is the
$r$ charge $q_{r}$ of the operators, defined as
\begin{equation}
\int dz\left\{ J_{r}\left(z\right)\mathcal{O}\left(y\right)\right\} =q_{r}\left(\mathcal{O}\right)\mathcal{O}\left(y\right).\label{eq:rcharge}
\end{equation}
The building blocks of $b{}_{n}$ satisfy:
\begin{equation}
\begin{array}{ccc}
\left\{ Q,-s^{\alpha}\partial\overline{\lambda}_{\alpha}\right\} =T_{\overline{\lambda}}, &  & \left\{ Q,G^{\alpha}\right\} =\left(\lambda^{\alpha},T_{\lambda}+T_{\textrm{matter}}\right),\\
\left[Q,H^{\alpha\beta}\right]=\left(\lambda^{[\alpha},G^{\beta]}\right), &  & \left\{ Q,K^{\alpha\beta\gamma}\right\} =\left(\lambda^{[\alpha},H^{\beta\gamma]}\right),\\
\left[Q,L^{\alpha\beta\gamma\lambda}\right]=\left(\lambda^{[\alpha},K^{\beta\gamma\lambda]}\right), &  & \left(\lambda^{[\alpha},L^{\beta\gamma\lambda\sigma]}\right)=0.
\end{array}
\end{equation}

Some observations should be made concerning the above operators:
\begin{itemize}
\item the ordering here, implemented through
\begin{equation}
\left(A,B\right)\left(y\right)\equiv\frac{1}{2\pi i}\oint\frac{dz}{z-y}A\left(z\right)B\left(y\right),\label{eq:ordering}
\end{equation}
plays a major role, allowing a correct manipulation of the quantum
corrections to the $b$ ghost. Obviously, a different ordering prescription
must not conflict with $\left\{ Q,b\right\} =T$.
\item the operator $O$ defined above is required because
\begin{equation}
\left(\frac{\left(\overline{\lambda}_{\alpha}r_{\beta}-\overline{\lambda}_{\beta}r_{\alpha}\right)\lambda^{\alpha}}{\left(\overline{\lambda}\lambda\right)^{2}},G^{\beta}\right)-\left(\frac{\overline{\lambda}_{\alpha}r_{\beta}}{\left(\overline{\lambda}\lambda\right)^{2}},\left(\lambda^{\alpha},G^{\beta}\right)-\left(\lambda^{\beta},G^{\alpha}\right)\right)\neq0,
\end{equation}
and
\begin{equation}
\left(\frac{\overline{\lambda}_{\alpha}}{\left(\overline{\lambda}\lambda\right)},\left(\lambda^{\alpha},T_{\lambda}\right)\right)-T_{\lambda}\neq0.
\end{equation}
One can see that $\left\{ Q,O\right\} $ precisely matches the above
inequalities. In \cite{Oda:2007ak}, besides \eqref{eq:ordering},
an alternative prescription was used, that conveniently absorbs the
operator $O$.
\item the quantum contribution to $G^{\alpha}$ is proportional to $\partial^{2}\theta^{\alpha}$.
The coefficient can be fixed by comparing the cubic pole in the OPE of
the energy-momentum tensor with both sides of the equation $\left\{ Q,G^{\alpha}\right\} =\left(\lambda^{\alpha},T\right)$,
or directly through the usual $U\left(5\right)$ decomposition%
\footnote{That explains why the coefficient used here differs from the one used in \cite{Berkovits:2004px,Berkovits:2005bt},
where $G^{\alpha}$ was required to be primary. Equation \eqref{eq:OPETGalpha}
shows that this is not the case, since $\left(\lambda^{\alpha},T\right)$
is not a primary field. 

\cite{Oda:2007ak} contains a detailed discussion on this subject. There, $G^{\alpha}$ and $\hat{G}^{\alpha}$ denote the primary and the non-primary constructions, respectively.%
}.
\end{itemize}
\

The last observation is directly related to the fact that the $b$
ghost is a conformal weight $2$ primary field \cite{Oda:2007ak},
\begin{equation}
T\left(z\right)b\left(y\right)\sim2\frac{b}{\left(z-y\right)^{2}}+\frac{\partial b}{\left(z-y\right)}.
\end{equation}
 This result is reproduced in appendix \ref{sec:Conventions}. Note
also that $b$ is a Lorentz scalar and manifestly supersymmetric.

Another interesting property of the $b$ ghost is the pole structure
of its OPE with the BRST current:

\begin{equation}
J_{BRST}\left(z\right)b\left(y\right)\sim\frac{3}{\left(z-y\right)^{3}}+\frac{J}{\left(z-y\right)^{2}}+\frac{T}{\left(z-y\right)},
\end{equation}
where
\begin{eqnarray}
J & = & J_{\lambda}+J_{r}-2\frac{\overline{\lambda}\partial\lambda}{\overline{\lambda}\lambda}+2\frac{r\partial\theta}{\overline{\lambda}\lambda}-2\frac{\left(r\lambda\right)\left(\overline{\lambda}\partial\theta\right)}{\left(\overline{\lambda}\lambda\right)^{2}}.\nonumber \\
 & = & J_{\lambda}-J_{\overline{\lambda}}-\left\{ Q,\left(S+2\frac{\overline{\lambda}\partial\theta}{\overline{\lambda}\lambda}\right)\right\} .
\end{eqnarray}
With a BRST transformation, the U$\left(1\right)$ current can be
brought into a more natural form, without changing the ghost numbers
of the BRST charge and the $b$ ghost.

To verify the interpretation of $J$ as the ghost number current,
it is worth noting that,
\begin{eqnarray}
T\left(z\right)J\left(y\right) & \sim & -\frac{3}{\left(z-y\right)^{3}}+\frac{J}{\left(z-y\right)^{2}}+\frac{\partial J}{\left(z-y\right)},\\
J\left(z\right)J_{BRST}\left(y\right) & \sim & \frac{J_{BRST}}{\left(z-y\right)},\\
J\left(z\right)b\left(y\right) & \sim & -\frac{b}{\left(z-y\right)}.
\end{eqnarray}

Together, $b$, $T$, $J_{BRST}$ and $J$ may describe a twisted
$\mathcal{N}=2$ $\hat{c}=3$ critical topological string \cite{Berkovits:2005bt}.
The untwisted version would satisfy
\[
\begin{array}{cc}
T'\left(z\right)T'\left(y\right)\sim\frac{\left(9/2\right)}{\left(z-y\right)^{4}}+2\frac{T'}{\left(z-y\right)^{2}}+\frac{\partial T'}{\left(z-y\right)}, & T'\left(z\right)J\left(y\right)\sim\frac{J}{\left(z-y\right)^{2}}+\frac{\partial J}{\left(z-y\right)},\end{array}
\]
\[
\begin{array}{cc}
T'\left(z\right)G^{+}\left(y\right)\sim\frac{3}{2}\frac{G^{+}}{\left(z-y\right)^{2}}+\frac{\partial G^{+}}{\left(z-y\right)}, & T'\left(z\right)G^{-}\left(y\right)\sim\frac{3}{2}\frac{G^{-}}{\left(z-y\right)^{2}}+\frac{\partial G^{-}}{\left(z-y\right)},\end{array}
\]
\[
\begin{array}{cc}
J\left(z\right)G^{+}\left(y\right)\sim\frac{G^{+}\left(y\right)}{\left(z-y\right)}, & J\left(z\right)G^{-}\left(y\right)\sim-\frac{G^{-}\left(y\right)}{\left(z-y\right)},\end{array}
\]
\[
\begin{array}{cc}
J\left(z\right)J\left(y\right)\sim\frac{3}{\left(z-y\right)^{2}}, & G^{+}\left(z\right)G^{-}\left(y\right)\sim\frac{3}{\left(z-y\right)^{3}}+\frac{J}{\left(z-y\right)^{2}}+\frac{T'+\frac{1}{2}\partial J}{\left(z-y\right)},\end{array}
\]
\[
\begin{array}{cc}
G^{+}\left(z\right)G^{+}\left(y\right)\sim\textrm{regular}, & G^{-}\left(z\right)G^{-}\left(y\right)\sim\textrm{regular},\end{array}
\]
 where $G^{+}=J_{BRST}$, $G^{-}=b$ and $T'=T-\frac{1}{2}\partial J$.
The twist here means $T'\to T'-\frac{1}{2}\partial J$, which modifies
the conformal weights of the ghosts $\lambda$ and $r$ from $\frac{1}{2}$
to $0$ and turns the central charge off.

By examining this set of OPE's, one notes that $b$ must be nilpotent
in order for the non-minimal pure spinor formalism to be viewed as
a topological string. This property will now be rigorously demonstrated.

\subsection{Nilpotency}

\

The OPE of the $b$ ghost with itself can be cast as
\begin{equation}
b\left(z\right)b\left(y\right)\sim\frac{O_{0}}{\left(z-y\right)^{4}}+\frac{O_{1}}{\left(z-y\right)^{3}}+\frac{O_{2}}{\left(z-y\right)^{2}}+\frac{O_{3}}{\left(z-y\right)},
\end{equation}
for there are no (covariant, supersymmetric) negative conformal weight
fields in the theory. Due to its anticommuting character, $b\left(z\right)b\left(y\right)=-b\left(y\right)b\left(z\right)$,
implying that
\begin{equation}
b\left(z\right)b\left(y\right)\sim\frac{O_{1}}{\left(z-y\right)^{3}}+\frac{1}{2}\frac{\partial O_{1}}{\left(z-y\right)^{2}}+\frac{O_{3}}{\left(z-y\right)}.\label{eq:OPEbbrestricted}
\end{equation}

Furthermore, since $\left\{ Q,b\right\} =T$ and $b$ is a primary
field of conformal weight $2$,
\begin{eqnarray}
\left\{ Q,b\left(z\right)\right\} b\left(y\right)-b\left(z\right)\left\{ Q,b\left(y\right)\right\}  & = & T\left(z\right)b\left(y\right)-b\left(z\right)T\left(y\right)\nonumber \\
 & \sim & \textrm{regular},
\end{eqnarray}
or, equivalently,
\begin{eqnarray}
\left\{ Q,b\left(z\right)b\left(y\right)\right\}  & \sim & \frac{\left\{ Q,O_{1}\right\} }{\left(z-y\right)^{3}}+\frac{1}{2}\frac{\partial\left\{ Q,O_{1}\right\} }{\left(z-y\right)^{2}}+\frac{\left\{ Q,O_{3}\right\} }{\left(z-y\right)}.
\end{eqnarray}
Comparing both expressions, one concludes that $O_{1}$ and $O_{3}$
are BRST closed.

Taking now into account the specific form of the $b$ ghost for the
non-minimal pure spinor formalism, given in \eqref{eq:quantumb},
it is a simple task to verify that the cubic poles are all proportional
to the constraints \eqref{eq:nonpsconstraint} and \eqref{eq:lambdabarrconstraint}.
The possible terms will be listed below:
\begin{itemize}
\item $b_{-1}$ may give rise to cubic poles only in the OPE with $b_{3}$,
due to ordering effects. The different terms are proportional to
\begin{equation}
\begin{array}{cc}
\left(\overline{\lambda}\gamma_{mnp}r\right)\left(\partial\overline{\lambda}\gamma^{pqr}r\right)\left(\overline{\lambda}\gamma^{mn}\lambda\right)\left(\overline{\lambda}\gamma_{qr}\lambda\right), & \left(\overline{\lambda}\gamma_{mnp}r\right)\left(\partial\overline{\lambda}\gamma^{pqr}r\right)\left(\overline{\lambda}\gamma^{mn}\gamma_{qr}\lambda\right),\\
\left(\overline{\lambda}\gamma_{mnp}\partial\overline{\lambda}\right)\left(r\gamma^{pqr}r\right)\left(\overline{\lambda}\gamma^{mn}\lambda\right)\left(\overline{\lambda}\gamma_{qr}\lambda\right), & \left(\overline{\lambda}\gamma_{mnp}\partial\overline{\lambda}\right)\left(r\gamma^{pqr}r\right)\left(\overline{\lambda}\gamma^{mn}\gamma_{qr}\lambda\right).
\end{array}
\end{equation}

\item $b_{0}$ has cubic poles with itself, $b_{1}$, $b_{2}$ and $b_{3}$:

\begin{itemize}
\item in $b_{0}\left(z\right)b_{0}\left(y\right)$, it comes from the multiple
contractions of $\Pi^{m}\left(\gamma_{m}d\right)^{\alpha}$ with itself
and from its single contraction with $\partial^{2}\theta^{\beta}$,
both proportional to $\Pi^{m}\left(\overline{\lambda}\gamma_{m}\overline{\lambda}\right)$.
\item for $b_{0}\left(z\right)b_{1}\left(y\right)$, it will arise in the
contractions of $\left(d\gamma^{mnp}d\right)\left(\overline{\lambda}\lambda\right)^{-2}$
with all the terms in $b_{0}$, being proportional to $\left(\overline{\lambda}\gamma_{mnp}r\right)\left(\overline{\lambda}\gamma^{mnp}d\right)$.
\item in the OPE $b_{0}\left(z\right)b_{2}\left(y\right)$, the multiple
contractions of $N^{mn}\left(\gamma^{p}d\right)^{\alpha}$ will give
cubic poles like:
\begin{equation}
\begin{array}{c}
\left(\overline{\lambda}\gamma_{mnp}r\right)N^{mn}\left(\overline{\lambda}\gamma^{p}r\right),\\
\left(\overline{\lambda}\gamma_{mnp}r\right)J\left(\overline{\lambda}\gamma^{mn}\lambda\right)\left(\overline{\lambda}\gamma^{p}r\right),\\
\partial\left[\left(\overline{\lambda}\gamma^{mn}\lambda\right)\overline{\lambda}_{\alpha}\right]\left(\overline{\lambda}\gamma_{mnp}r\right)\left(\gamma^{p}r\right)^{\alpha}.
\end{array}
\end{equation}

\item finally, in $b_{0}\left(z\right)b_{3}\left(y\right)$, the cubic poles
are of the form:
\begin{equation}
\begin{array}{c}
\left(\overline{\lambda}\partial\theta\right)\left(\overline{\lambda}\gamma^{mn}\lambda\right)\left(\overline{\lambda}\gamma_{qr}\lambda\right)\left(\overline{\lambda}\gamma_{mnp}r\right)\left(r\gamma^{pqr}r\right),\\
\left(\overline{\lambda}\gamma^{mn}\partial\theta\right)\left(\overline{\lambda}\gamma_{qr}\lambda\right)\left(\overline{\lambda}\gamma_{mnp}r\right)\left(r\gamma^{pqr}r\right).
\end{array}
\end{equation}

\end{itemize}
\item $b_{1}$ has cubic poles with itself, $b_{2}$ and $b_{3}$:

\begin{itemize}
\item in $b_{1}\left(z\right)b_{1}\left(y\right)$, they are of the form
\begin{equation}
\begin{array}{c}
\partial\left(\overline{\lambda}\gamma_{mnp}r\right)\left(\overline{\lambda}\gamma^{mnp}r\right),\\
\left(\overline{\lambda}\gamma_{mnp}r\right)\left(\overline{\lambda}\gamma^{mnq}r\right)N_{\phantom{p}q}^{p},\\
\partial\left(\overline{\lambda}\gamma_{mnp}r\right)\left(\overline{\lambda}\gamma^{pqr}r\right)\left(\overline{\lambda}\gamma^{mn}\lambda\right)\left(\overline{\lambda}\gamma_{qr}\lambda\right).
\end{array}
\end{equation}

\item for $b_{1}\left(z\right)b_{2}\left(y\right)$, the only possible cubic
poles are proportional to
\begin{equation}
\begin{array}{c}
\left(\overline{\lambda}\gamma_{mnp}r\right)\left(\overline{\lambda}\gamma_{qrs}r\right)\left(\overline{\lambda}\gamma^{mn}\lambda\right)\left(r\gamma^{qrs}\gamma^{p}\partial\theta\right),\\
\left(\overline{\lambda}\gamma_{mnp}r\right)\left(\overline{\lambda}\gamma^{mnq}r\right)\left(r\gamma_{q}\gamma^{p}\partial\theta\right).
\end{array}
\end{equation}

\item the cubic poles arising in $b_{1}\left(z\right)b_{3}\left(y\right)$
come from the multiple contractions of $N^{mn}\Pi^{p}\left(\overline{\lambda}\lambda\right)^{-2}$
with $b_{3}$, and are given by
\begin{equation}
\begin{array}{c}
\left(\overline{\lambda}\gamma_{mnp}r\right)\left(\overline{\lambda}\gamma^{mnq}r\right)\left(r\gamma_{qrs}r\right)\left(\overline{\lambda}\gamma^{rs}\lambda\right)\Pi^{p},\\
\left(\overline{\lambda}\gamma_{mnp}r\right)\left(\overline{\lambda}\gamma^{mn}\lambda\right)\Pi^{p}\left(\overline{\lambda}\gamma_{qrs}r\right)\left(\overline{\lambda}\gamma^{qr}\lambda\right)\left(r\gamma^{stu}r\right)\left(\overline{\lambda}\gamma_{tu}\lambda\right),\\
\left(\overline{\lambda}\gamma_{mnp}r\right)\left(\overline{\lambda}\gamma^{mn}\lambda\right)\Pi^{p}\left(\overline{\lambda}\gamma_{qrs}r\right)\left(\overline{\lambda}\gamma^{qr}\gamma_{tu}\lambda\right)\left(r\gamma^{stu}r\right).
\end{array}
\end{equation}

\end{itemize}
\item $b_{2}$ has cubic poles with itself and with $b_{3}$:

\begin{itemize}
\item in $b_{2}\left(z\right)b_{2}\left(y\right)$, they are of the form
\begin{equation}
\begin{array}{c}
\left(\overline{\lambda}\gamma_{mnp}r\right)\left(\overline{\lambda}\gamma_{qrs}r\right)\left(r\gamma^{pst}r\right)\Pi_{t}\eta^{mq}\eta^{nr},\\
\left(\overline{\lambda}\gamma_{mnp}r\right)\left(\overline{\lambda}\gamma^{mn}\lambda\right)\left(\overline{\lambda}\gamma_{qrs}r\right)\left(\overline{\lambda}\gamma^{qr}\lambda\right)\left(r\gamma^{pst}r\right)\Pi_{t}.
\end{array}
\end{equation}

\item for $b_{2}\left(z\right)b_{3}\left(y\right)$, $d_{\alpha}$ appearing
in $b_{2}$ is inert and there are only contractions involving the
ghost Lorentz currents:
\begin{equation}
\begin{array}{c}
\left(\overline{\lambda}\gamma_{mnp}r\right)\left(r\gamma^{p}d\right)\left(\overline{\lambda}\gamma^{mnq}r\right)\left(r\gamma_{qrs}r\right)\left(\overline{\lambda}\gamma^{rs}\lambda\right),\\
\left(\overline{\lambda}\gamma_{mnp}r\right)\left(\overline{\lambda}\gamma^{mn}\lambda\right)\left(r\gamma^{p}d\right)\left(\overline{\lambda}\gamma_{qrs}r\right)\left(\overline{\lambda}\gamma^{qr}\lambda\right)\left(r\gamma^{stu}r\right)\left(\overline{\lambda}\gamma_{tu}\lambda\right).
\end{array}
\end{equation}

\end{itemize}
\item the cubic poles of $b_{3}\left(z\right)b_{3}\left(y\right)$ involve
all possible contractions of the the Lorentz generators and will give
similar results to the ones above, only with more $r$'s.
\end{itemize}
Due to the pure spinor constraints,
\begin{equation}
\left(\overline{\lambda}\gamma^{mn}\right)_{\alpha}\left(\overline{\lambda}\gamma_{mnp}r\right)=\left(\overline{\lambda}\gamma^{mn}\right)_{\alpha}\left(\overline{\lambda}\gamma_{mnp}\partial\overline{\lambda}\right)=\left(\overline{\lambda}\gamma_{mnp}r\right)\left(\overline{\lambda}\gamma^{mnp}\right)^{\alpha}=0,\label{eq:cubicpolevanishing}
\end{equation}
and every expression listed contains at least one of these types of
contractions. Consequently, $O_{1}=0$ and
\begin{equation}
b\left(z\right)b\left(y\right)\sim\frac{O_{3}}{\left(z-y\right)}.\label{eq:opebbsimp}
\end{equation}

It is clear from \eqref{eq:quantumb}, that $O_{3}$ can only be composed
with supersymmetric invariants: matter fields ($\Pi^{m}$, $d_{\alpha}$,
$\partial\theta^{\alpha}$); ghost currents from the minimal sector
($N^{mn}$, $J$); ghost fields ($\lambda^{\alpha}$,$\overline{\lambda}_{\alpha}$,$r_{\alpha}$);
and, in principle, their partial derivatives.

In \cite{Chandia:2010ix}, the vanishing of $O_{3}$ has been argued
as follows. The author assumed that all partial derivatives of $r_{\alpha}$
that may appear in the OPE \eqref{eq:opebbsimp} can be removed due
to the pure spinor constraint, since
\begin{equation}
\overline{\lambda}\gamma^{m}\partial r=-\partial\overline{\lambda}\gamma^{m}r.
\end{equation}
Based on that assumption, all the $r_{\alpha}$ dependence of $O_{3}$
could be made explicitly through
\begin{equation}
O_{3}=\Omega+r_{\alpha}\Omega^{\alpha}+r_{\alpha}r_{\beta}\Omega^{\alpha\beta}+\ldots
\end{equation}
where the $\Omega$'s are supersymmetric, ghost number $-2$, conformal
weight $3$, BRST closed operators. Since the BRST charge can be split
into two pieces according to the $r$-charge\begin{subequations}
\begin{eqnarray}
Q & = & Q_{0}+Q_{1},\\
Q_{0} & = & \oint\left(\lambda^{\alpha}d_{\alpha}\right),\\
Q_{1} & = & \oint\left(\overline{\omega}^{\alpha}r_{\alpha}\right),
\end{eqnarray}
\end{subequations}requiring $\left[Q,O_{3}\right]=0$, implies $\left[Q_{0},\Omega\right]=0$.
Then, it has been shown that there are no $\Omega$ with the above
requisites satisfying $\left[Q_{0},\Omega\right]=0$, so it vanishes
identically. Then, $\Omega=0$ implies $\left[Q_{0},\Omega^{\alpha}\right]=0$.
Again, this can be demonstrated to vanish. Pursuing this argument,
the nilpotency of the $b$ ghost was obtained in \cite{Chandia:2010ix}.

However, the absence of $\partial^{n}r_{\alpha}$ in $O_{3}$ is incorrect,
as will be illustrated soon, which means that the cohomology argument
of \cite{Chandia:2010ix}, summarized above, must be extended, as
will now be done.

The computation of \eqref{eq:opebbsimp} is organized according to
the $r$-charge of the operators, that is
\begin{equation}
O_{3}=\left(bb\right)_{0}+\left(bb\right)_{1}+\left(bb\right)_{2}+\left(bb\right)_{3}+\left(bb\right)_{4}+\left(bb\right)_{5}+\left(bb\right)_{6}.
\end{equation}
To make the expressions more clear, the ordering notation will be
dropped and $\alpha'$ will be set to $2$.

The first term, $\left(bb\right)_{0}$, is given by
\begin{eqnarray}
\left(bb\right)_{0} & \equiv & \int dz\left\{ b_{0}\left(z\right)b_{0}\left(y\right)+b_{-1}\left(z\right)b_{1}\left(y\right)+b_{1}\left(z\right)b_{-1}\left(y\right)\right\} \\
 & = & \alpha_{01}\frac{N^{mn}\left(\overline{\lambda}\gamma_{mn}\partial\theta\right)\left(\overline{\lambda}\partial\theta\right)}{\left(\overline{\lambda}\lambda\right)^{2}}+\alpha_{02}\frac{\left(\overline{\lambda}\gamma_{mnp}\partial\overline{\lambda}\right)N^{mn}\Pi^{p}}{\left(\overline{\lambda}\lambda\right)^{2}}\nonumber \\
 & + & \alpha_{03}\frac{\Pi^{m}\left(\overline{\lambda}\partial\theta\right)\left(\overline{\lambda}\gamma_{m}d\right)}{\left(\overline{\lambda}\lambda\right)^{2}}+\alpha_{04}\frac{\left(\overline{\lambda}\gamma_{mnp}\partial\overline{\lambda}\right)\left(d\gamma^{mnp}d\right)}{\left(\overline{\lambda}\lambda\right)^{2}}\nonumber \\
 & + & \alpha_{05}\frac{\Pi^{m}\left(\overline{\lambda}\gamma_{m}\partial^{2}\overline{\lambda}\right)}{\left(\overline{\lambda}\lambda\right)^{2}}+\alpha_{06}\frac{\left(\overline{\lambda}\partial\theta\right)\left(\overline{\lambda}\partial^{2}\theta\right)}{\left(\overline{\lambda}\lambda\right)^{2}}+\alpha_{07}\frac{\left(\overline{\lambda}\partial\theta\right)\left(\partial\overline{\lambda}\partial\theta\right)}{\left(\overline{\lambda}\lambda\right)^{2}},
\end{eqnarray}
where $\alpha_{0n}$ are just numerical coefficients. By a direct
computation, it is relatively simple to show the vanishing of $\left(bb\right)_{0}$.
It is enough to compute $\left[Q,\left(bb\right)_{0}\right]$ and
use the BRST argument mentioned above. Note that $\left[Q,O_{3}\right]=0$
implies the vanishing of
\begin{eqnarray*}
\left[Q_{0},\left(bb\right)_{0}\right] & = & \alpha_{01}\frac{N^{mn}\left(\overline{\lambda}\gamma_{mn}\partial\lambda\right)\left(\overline{\lambda}\partial\theta\right)}{\left(\overline{\lambda}\lambda\right)^{2}}-\alpha_{01}\frac{\frac{1}{2}\left(d\gamma^{mn}\lambda\right)\left(\overline{\lambda}\gamma_{mn}\partial\theta\right)\left(\overline{\lambda}\partial\theta\right)}{\left(\overline{\lambda}\lambda\right)^{2}}\\
 & - & \alpha_{01}\frac{N^{mn}\left(\overline{\lambda}\gamma_{mn}\partial\theta\right)\left(\overline{\lambda}\partial\lambda\right)}{\left(\overline{\lambda}\lambda\right)^{2}}-\alpha_{02}\frac{\left(\overline{\lambda}\gamma_{mnp}\partial\overline{\lambda}\right)\frac{1}{2}\left(d\gamma^{mn}\lambda\right)\Pi^{p}}{\left(\overline{\lambda}\lambda\right)^{2}}\\
 & + & \alpha_{02}\frac{\left(\overline{\lambda}\gamma_{mnp}\partial\overline{\lambda}\right)N^{mn}\left(\lambda\gamma^{p}\partial\theta\right)}{\left(\overline{\lambda}\lambda\right)^{2}}-\alpha_{03}\frac{\Pi^{m}\left(\overline{\lambda}\partial\theta\right)\left(\overline{\lambda}\gamma_{m}\gamma_{n}\partial\theta\right)\Pi^{n}}{\left(\overline{\lambda}\lambda\right)^{2}}\\
 & + & \alpha_{03}\frac{\Pi^{m}\left(\overline{\lambda}\partial\lambda\right)\left(\overline{\lambda}\gamma_{m}d\right)}{\left(\overline{\lambda}\lambda\right)^{2}}+\alpha_{03}\frac{\left(\lambda\gamma^{m}\partial\theta\right)\left(\overline{\lambda}\partial\theta\right)\left(\overline{\lambda}\gamma_{m}d\right)}{\left(\overline{\lambda}\lambda\right)^{2}}\\
 & + & \alpha_{04}\frac{2\left(\overline{\lambda}\gamma_{mnp}\partial\overline{\lambda}\right)\left(d\gamma^{mnp}\gamma^{q}\lambda\right)\Pi_{q}}{\left(\overline{\lambda}\lambda\right)^{2}}+\alpha_{06}\frac{\left(\overline{\lambda}\partial\lambda\right)\left(\overline{\lambda}\partial^{2}\theta\right)}{\left(\overline{\lambda}\lambda\right)^{2}}-\alpha_{06}\frac{\left(\overline{\lambda}\partial\theta\right)\left(\overline{\lambda}\partial^{2}\lambda\right)}{\left(\overline{\lambda}\lambda\right)^{2}}\\
 & + & \alpha_{05}\frac{\left(\lambda\gamma^{m}\partial\theta\right)\left(\overline{\lambda}\gamma_{m}\partial^{2}\overline{\lambda}\right)}{\left(\overline{\lambda}\lambda\right)^{2}}+\alpha_{07}\frac{\left(\overline{\lambda}\partial\lambda\right)\left(\partial\overline{\lambda}\partial\theta\right)}{\left(\overline{\lambda}\lambda\right)^{2}}-\alpha_{07}\frac{\left(\overline{\lambda}\partial\theta\right)\left(\partial\overline{\lambda}\partial\lambda\right)}{\left(\overline{\lambda}\lambda\right)^{2}}.
\end{eqnarray*}
The Lorentz generators $N^{mn}$ appear in three terms. It is straightforward
to check that they are not related by a Fierz decomposition of the
spinors, implying that $\alpha_{01}=\alpha_{02}=0$. Now, there is
only one term that contributes with one $d_{\alpha}$ and two $\partial\theta^{\alpha}$,
so $\alpha_{03}=0$, which, on the other hand, imply that $\alpha_{04}=0$,
since the term with one $d_{\alpha}$ and one $\Pi^{m}$ cannot be
cancelled anymore. The vanishing of $\alpha_{05}$, $\alpha_{06}$
and $\alpha_{07}$ is evident, since they do not possibly cancel each
other. \emph{There is no linear combination of the above operators
that can be annihilated by} $Q_{0}$, \emph{therefore $\left(bb\right)_{0}=0$}.

The second term, $\left(bb\right)_{1}$, is
\begin{eqnarray}
\left(bb\right)_{1} & \equiv & \int dz\left\{ b_{0}\left(z\right)b_{1}\left(y\right)+b_{1}\left(z\right)b_{0}\left(y\right)+b_{-1}\left(z\right)b_{2}\left(y\right)+b_{2}\left(z\right)b_{-1}\left(y\right)\right\} \\
 & = & \alpha_{11}\frac{\left(\overline{\lambda}\gamma_{mnp}r\right)N^{mn}\Pi^{p}\left(\overline{\lambda}\partial\theta\right)}{\left(\overline{\lambda}\lambda\right)^{3}}+\alpha_{12}\frac{\left(\overline{\lambda}\gamma_{mnp}r\right)N^{mn}\left(\partial\overline{\lambda}\gamma^{p}d\right)}{\left(\overline{\lambda}\lambda\right)^{3}}+\nonumber \\
 & + & \alpha_{13}\frac{\left(\overline{\lambda}\gamma_{mnp}\partial\overline{\lambda}\right)N^{mn}\left(r\gamma^{p}d\right)}{\left(\overline{\lambda}\lambda\right)^{3}}+\alpha_{14}\frac{\left(\overline{\lambda}\gamma_{mnp}r\right)\left(d\gamma^{mnp}d\right)\left(\overline{\lambda}\partial\theta\right)}{\left(\overline{\lambda}\lambda\right)^{3}}\nonumber \\
 & + & \alpha_{15}\frac{\left(\overline{\lambda}\gamma_{mnp}r\right)\Pi^{m}\left(\partial\overline{\lambda}\gamma^{np}\partial\theta\right)}{\left(\overline{\lambda}\lambda\right)^{3}}+\alpha_{16}\frac{\left(\overline{\lambda}\gamma_{m}\partial^{2}\overline{\lambda}\right)\left(r\gamma^{m}d\right)}{\left(\overline{\lambda}\lambda\right)^{3}}.
\end{eqnarray}
Since $\left[Q_{1},\left(bb\right)_{0}\right]=0$, $\left[Q_{0},\left(bb\right)_{1}\right]$
must also vanish:
\begin{eqnarray*}
\left[Q_{0},\left(bb\right)_{1}\right] & = & \alpha_{11}\frac{\frac{1}{2}\left(\overline{\lambda}\gamma_{mnp}r\right)\left(d\gamma^{mn}\lambda\right)\Pi^{p}\left(\overline{\lambda}\partial\theta\right)}{\left(\overline{\lambda}\lambda\right)^{3}}-\alpha_{11}\frac{\left(\overline{\lambda}\gamma_{mnp}r\right)N^{mn}\left(\lambda\gamma^{p}\partial\theta\right)\left(\overline{\lambda}\partial\theta\right)}{\left(\overline{\lambda}\lambda\right)^{3}}\\
 & - & \alpha_{11}\frac{\left(\overline{\lambda}\gamma_{mnp}r\right)N^{mn}\Pi^{p}\left(\overline{\lambda}\partial\lambda\right)}{\left(\overline{\lambda}\lambda\right)^{3}}+\alpha_{12}\frac{\frac{1}{2}\left(\overline{\lambda}\gamma_{mnp}r\right)\left(d\gamma^{mn}\lambda\right)\left(\partial\overline{\lambda}\gamma^{p}d\right)}{\left(\overline{\lambda}\lambda\right)^{3}}\\
 & + & \alpha_{12}\frac{\left(\overline{\lambda}\gamma_{mnp}r\right)N^{mn}\left(\partial\overline{\lambda}\gamma^{p}\gamma^{q}\lambda\right)\Pi_{q}}{\left(\overline{\lambda}\lambda\right)^{3}}-\alpha_{13}\frac{\frac{1}{2}\left(\overline{\lambda}\gamma_{mnp}\partial\overline{\lambda}\right)\left(d\gamma^{mn}\lambda\right)\left(r\gamma^{p}d\right)}{\left(\overline{\lambda}\lambda\right)^{3}}\\
 & + & \alpha_{13}\frac{\left(\overline{\lambda}\gamma_{mnp}\partial\overline{\lambda}\right)N^{mn}\left(r\gamma^{p}\gamma^{q}\lambda\right)\Pi_{q}}{\left(\overline{\lambda}\lambda\right)^{3}}-\alpha_{14}\frac{2\left(\overline{\lambda}\gamma_{mnp}r\right)\left(d\gamma^{mnp}\gamma^{q}\lambda\right)\Pi_{q}\left(\overline{\lambda}\partial\theta\right)}{\left(\overline{\lambda}\lambda\right)^{3}}\\
 & - & \alpha_{14}\frac{\left(\overline{\lambda}\gamma_{mnp}r\right)\left(d\gamma^{mnp}d\right)\left(\overline{\lambda}\partial\lambda\right)}{\left(\overline{\lambda}\lambda\right)^{3}}-\alpha_{15}\frac{\left(\overline{\lambda}\gamma_{mnp}r\right)\left(\lambda\gamma^{m}\partial\theta\right)\left(\partial\overline{\lambda}\gamma^{np}\partial\theta\right)}{\left(\overline{\lambda}\lambda\right)^{3}}\\
 & - & \alpha_{15}\frac{\left(\overline{\lambda}\gamma_{mnp}r\right)\Pi^{m}\left(\partial\overline{\lambda}\gamma^{np}\partial\lambda\right)}{\left(\overline{\lambda}\lambda\right)^{3}}+\alpha_{16}\frac{\left(\overline{\lambda}\gamma_{m}\partial^{2}\overline{\lambda}\right)\left(r\gamma^{m}\gamma^{n}\lambda\right)\Pi_{n}}{\left(\overline{\lambda}\lambda\right)^{3}}.
\end{eqnarray*}
There is only one term that contains one Lorentz generator $N^{mn}$
and two $\partial\theta^{\alpha}$, so $\alpha_{11}=0$. Now, there
are two other terms that contain $N^{mn}$, but they are unrelated
to any Fierz decomposition, implying that $\alpha_{12}=\alpha_{13}=0$.
The remaining terms are obviously independent: $\alpha_{14}=0$, since
it is the only one with $\left(d\gamma^{mnp}d\right)$; $\alpha_{15}=0$,
as no other term contains two $\partial\theta^{\alpha}$; and $\alpha_{16}=0$,
for there is nothing else to cancel it. As $\left(bb\right)_{0}$,
$\left(bb\right)_{1}$ is not BRST closed for any set of coefficients
$\alpha_{1n}$ and $\left(bb\right)_{1}=0$ is the single possibility
left.

Going on,
\begin{equation}
\left(bb\right)_{2}\equiv\int dz\left\{ b_{0}\left(z\right)b_{2}\left(y\right)+b_{2}\left(z\right)b_{0}\left(y\right)+b_{1}\left(z\right)b_{1}\left(y\right)+b_{-1}\left(z\right)b_{3}\left(y\right)+b_{3}\left(z\right)b_{-1}\left(y\right)\right\}
\end{equation}
can be written as

\begin{eqnarray}
\left(bb\right)_{2} & = & \alpha_{21}\frac{\left(\overline{\lambda}\gamma_{mnp}r\right)\left(r\gamma^{p}d\right)N^{mn}\left(\overline{\lambda}\partial\theta\right)}{\left(\overline{\lambda}\lambda\right)^{4}}+\alpha_{22}\frac{\left(\overline{\lambda}\gamma_{m}\partial r\right)\left(r\gamma^{m}d\right)\left(\overline{\lambda}\partial\theta\right)}{\left(\overline{\lambda}\lambda\right)^{4}}\nonumber \\
 & + & \alpha_{23}\frac{\left(\overline{\lambda}\gamma_{mnp}r\right)\left(\partial\overline{\lambda}\gamma^{pqr}r\right)N^{mn}N_{qr}}{\left(\overline{\lambda}\lambda\right)^{4}}+\alpha_{24}\frac{\left(\overline{\lambda}\gamma_{m}\partial r\right)\left(\overline{\lambda}\gamma_{n}d\right)\left(r\gamma^{mn}\partial\theta\right)}{\left(\overline{\lambda}\lambda\right)^{4}}\nonumber \\
 & + & \alpha_{25}\frac{\left(\overline{\lambda}\gamma_{mnp}r\right)\left(\partial\overline{\lambda}\gamma_{q}r\right)N^{mn}N^{pq}}{\left(\overline{\lambda}\lambda\right)^{4}}+\alpha_{26}\frac{\left(\overline{\lambda}\gamma_{mnp}r\right)\left(r\gamma^{p}\partial^{2}\overline{\lambda}\right)N^{mn}}{\left(\overline{\lambda}\lambda\right)^{4}}\nonumber \\
 & + & \alpha_{27}\frac{\left(\overline{\lambda}\gamma_{m}\partial r\right)\left(r\gamma^{m}\partial^{2}\overline{\lambda}\right)}{\left(\overline{\lambda}\lambda\right)^{4}}+\alpha_{28}\frac{\left(r\gamma_{m}\partial^{2}\overline{\lambda}\right)\left(\overline{\lambda}\gamma^{m}\partial r\right)}{\left(\overline{\lambda}\lambda\right)^{4}}.
\end{eqnarray}
The last line of the expression is $Q_{0}$-closed. In computing $\left[Q_{0},\left(bb\right)_{2}\right]$,
\begin{eqnarray*}
\left[Q_{0},\left(bb\right)_{2}\right] & = & \alpha_{21}\frac{\frac{1}{2}\left(\overline{\lambda}\gamma_{mnp}r\right)\left(r\gamma^{p}d\right)\left(d\gamma^{mn}\lambda\right)\left(\overline{\lambda}\partial\theta\right)}{\left(\overline{\lambda}\lambda\right)^{4}}-\alpha_{21}\frac{\left(\overline{\lambda}\gamma_{mnp}r\right)\left(r\gamma^{p}d\right)N^{mn}\left(\overline{\lambda}\partial\lambda\right)}{\left(\overline{\lambda}\lambda\right)^{4}}\\
 & - & \alpha_{21}\frac{\left(\overline{\lambda}\gamma_{mnp}r\right)\left(r\gamma^{p}\gamma^{q}\lambda\right)\Pi_{q}N^{mn}\left(\overline{\lambda}\partial\theta\right)}{\left(\overline{\lambda}\lambda\right)^{4}}-\alpha_{22}\frac{\left(\overline{\lambda}\gamma_{m}\partial r\right)\left(r\gamma^{m}\gamma^{n}\lambda\right)\Pi_{n}\left(\overline{\lambda}\partial\theta\right)}{\left(\overline{\lambda}\lambda\right)^{4}}\\
 & - & \alpha_{22}\frac{\left(\overline{\lambda}\gamma_{m}\partial r\right)\left(r\gamma^{m}d\right)\left(\overline{\lambda}\partial\lambda\right)}{\left(\overline{\lambda}\lambda\right)^{4}}-\alpha_{23}\frac{\frac{1}{2}\left(\overline{\lambda}\gamma_{mnp}r\right)\left(\partial\overline{\lambda}\gamma^{pqr}r\right)\left(d\gamma^{mn}\lambda\right)N_{qr}}{\left(\overline{\lambda}\lambda\right)^{4}}\\
 & - & \alpha_{23}\frac{\frac{1}{2}\left(\overline{\lambda}\gamma_{mnp}r\right)\left(\partial\overline{\lambda}\gamma^{pqr}r\right)N^{mn}\left(d\gamma_{qr}\lambda\right)}{\left(\overline{\lambda}\lambda\right)^{4}}-\alpha_{24}\frac{\left(\overline{\lambda}\gamma_{m}\partial r\right)\left(\overline{\lambda}\gamma_{n}d\right)\left(r\gamma^{mn}\partial\lambda\right)}{\left(\overline{\lambda}\lambda\right)^{4}}\\
 & + & \alpha_{24}\frac{\left(\overline{\lambda}\gamma_{m}\partial r\right)\left(\overline{\lambda}\gamma_{n}\gamma_{p}\lambda\right)\Pi^{p}\left(r\gamma^{mn}\partial\theta\right)}{\left(\overline{\lambda}\lambda\right)^{4}}-\alpha_{25}\frac{\frac{1}{2}\left(\overline{\lambda}\gamma_{mnp}r\right)\left(\partial\overline{\lambda}\gamma_{q}r\right)N^{mn}\left(d\gamma^{pq}\lambda\right)}{\left(\overline{\lambda}\lambda\right)^{4}}\\
 & - & \alpha_{25}\frac{\frac{1}{2}\left(\overline{\lambda}\gamma_{mnp}r\right)\left(\partial\overline{\lambda}\gamma_{q}r\right)\left(d\gamma^{mn}\lambda\right)N^{pq}}{\left(\overline{\lambda}\lambda\right)^{4}}-\alpha_{26}\frac{\frac{1}{2}\left(\overline{\lambda}\gamma_{mnp}r\right)\left(r\gamma^{p}\partial^{2}\overline{\lambda}\right)\left(d\gamma^{mn}\lambda\right)}{\left(\overline{\lambda}\lambda\right)^{4}},
\end{eqnarray*}
the terms that contain matter fields or the Lorentz current do not
vanish for any set $\alpha_{2n}$ of coefficients: $\alpha_{21}=0$,
for it is the single term that contains $N^{mn}$ and $\Pi^{m}$;
$\alpha_{22}=\alpha_{24}=0$, since they are the only ones that contribute
with one $\Pi^{m}$ and one $\partial\theta^{\alpha}$, but independently;
$\alpha_{23}=\alpha_{25}=0$, because they are the remaining (and
also independent) terms containing the Lorentz generator; and $\alpha_{26}=0$,
for it is not BRST closed.

$\left(bb\right)_{3}$ can be cast as:

\begin{eqnarray}
\left(bb\right)_{3} & \equiv & \int dz\left\{ b_{0}\left(z\right)b_{3}\left(y\right)+b_{3}\left(z\right)b_{0}\left(y\right)+b_{1}\left(z\right)b_{2}\left(y\right)+b_{2}\left(z\right)b_{1}\left(y\right)\right\} \\
 & = & \alpha_{31}\frac{\left(\overline{\lambda}\gamma_{mnp}r\right)\left(r\gamma^{pqr}r\right)N^{mn}N_{qr}\left(\overline{\lambda}\partial\theta\right)}{\left(\overline{\lambda}\lambda\right)^{5}}+\alpha_{32}\frac{\left(r\gamma_{mnp}r\right)\left(\overline{\lambda}\gamma^{p}\partial r\right)N^{mn}\left(\overline{\lambda}\partial\theta\right)}{\left(\overline{\lambda}\lambda\right)^{5}}\nonumber \\
 & + & \alpha_{33}\frac{\left(\overline{\lambda}\gamma_{m}\partial r\right)\left(r\gamma^{m}\partial r\right)\left(\overline{\lambda}\partial\theta\right)}{\left(\overline{\lambda}\lambda\right)^{5}}+\alpha_{34}\frac{\left(\overline{\lambda}\gamma_{m}\partial r\right)\left(\overline{\lambda}\gamma_{n}\partial r\right)\left(r\gamma^{mn}\partial\theta\right)}{\left(\overline{\lambda}\lambda\right)^{5}}\nonumber \\
 & + & \alpha_{35}\frac{\left(\overline{\lambda}\gamma_{m}\partial r\right)\left(\overline{\lambda}\gamma_{n}\partial r\right)\left(r\gamma^{mn}\lambda\right)\left(\overline{\lambda}\partial\theta\right)}{\left(\overline{\lambda}\lambda\right)^{6}}.
\end{eqnarray}
It is straightforward to see that the first two terms are not BRST
closed. One of the contributions of the first one contains two Lorentz
generators, that cannot be cancelled, so $\alpha_{31}=0$. The same
happens for the second one, which has a contribution in $\left[Q_{0},\left(bb\right)_{3}\right]$
with one Lorentz generator, not balanced by any other, thus $\alpha_{32}=0$.
The result of the computation of $\left[Q_{1},\left(bb\right)_{2}\right]+\left[Q_{0},\left(bb\right)_{3}\right]$
with the remaining terms is
\begin{eqnarray*}
\left[Q_{1},\left(bb\right)_{2}\right]+\left[Q_{0},\left(bb\right)_{3}\right] & = & \alpha_{27}\frac{4\left(\overline{\lambda}\gamma_{m}\partial r\right)\left(r\gamma^{m}\partial^{2}\overline{\lambda}\right)\left(r\lambda\right)}{\left(\overline{\lambda}\lambda\right)^{5}}-\alpha_{27}\frac{\left(r\gamma_{m}\partial r\right)\left(r\gamma^{m}\partial^{2}\overline{\lambda}\right)}{\left(\overline{\lambda}\lambda\right)^{4}}\\
 & - & \alpha_{27}\frac{\left(\overline{\lambda}\gamma_{m}\partial r\right)\left(r\gamma^{m}\partial^{2}r\right)}{\left(\overline{\lambda}\lambda\right)^{4}}+\alpha_{28}\frac{4\left(\overline{\lambda}\gamma_{m}\partial^{2}\overline{\lambda}\right)\left(r\gamma^{m}\partial r\right)\left(r\lambda\right)}{\left(\overline{\lambda}\lambda\right)^{5}}\\
 & - & \alpha_{28}\frac{\left(r\gamma_{m}\partial^{2}\overline{\lambda}\right)\left(r\gamma^{m}\partial r\right)}{\left(\overline{\lambda}\lambda\right)^{4}}-\alpha_{34}\frac{\left(\overline{\lambda}\gamma_{m}\partial r\right)\left(\overline{\lambda}\gamma_{n}\partial r\right)\left(r\gamma^{mn}\partial\lambda\right)}{\left(\overline{\lambda}\lambda\right)^{5}}\\
 & - & \alpha_{28}\frac{\left(\overline{\lambda}\gamma_{m}\partial^{2}r\right)\left(r\gamma^{m}\partial r\right)}{\left(\overline{\lambda}\lambda\right)^{4}}-\alpha_{33}\frac{\left(\overline{\lambda}\gamma_{m}\partial r\right)\left(r\gamma^{m}\partial r\right)\left(\overline{\lambda}\partial\lambda\right)}{\left(\overline{\lambda}\lambda\right)^{5}}\\
 & - & \alpha_{35}\frac{\left(\overline{\lambda}\gamma_{m}\partial r\right)\left(\overline{\lambda}\gamma_{n}\partial r\right)\left(r\gamma^{mn}\lambda\right)\left(\overline{\lambda}\partial\lambda\right)}{\left(\overline{\lambda}\lambda\right)^{6}}.
\end{eqnarray*}
Obviously, there is no nontrivial solution to $\left\{ \alpha_{27},\alpha_{28},\alpha_{33},\alpha_{34},\alpha_{35}\right\} $
that may lead to the vanishing of this equation, thus $\left(bb\right)_{2}=\left(bb\right)_{3}=0$.
Note that
\begin{equation}
\frac{\left(\overline{\lambda}\gamma_{m}\partial r\right)\left(r\gamma^{m}\partial r\right)\left(\overline{\lambda}\partial\theta\right)}{\left(\overline{\lambda}\lambda\right)^{5}}
\end{equation}
does not allow the removal of partial derivatives acting on $r$,
which contradicts the assumption of \cite{Chandia:2010ix}.

So far, the pure spinor constraints only have been used to reduce
the number of independent terms in the OPE computation. It turns out
that for $\left(bb\right)_{4}$, $\left(bb\right)_{5}$ and $\left(bb\right)_{6}$,
all possible terms being generated vanish due to the constraints.

For
\begin{equation}
\left(bb\right)_{4}\equiv\int dz\left\{ b_{1}\left(z\right)b_{3}\left(y\right)+b_{3}\left(z\right)b_{1}\left(y\right)+b_{2}\left(z\right)b_{2}\left(y\right)\right\} ,
\end{equation}
the simple poles are given by:
\begin{itemize}
\item terms with two $N$'s and one $\Pi$, like
\begin{equation}
\frac{\left(\overline{\lambda}\gamma_{mnp}r\right)\left(\overline{\lambda}\gamma_{qrs}r\right)\left(r\gamma^{pqt}r\right)N^{mn}N_{\phantom{r}t}^{r}\Pi^{s}}{\left(\overline{\lambda}\lambda\right)^{6}}.
\end{equation}
Since $\left(r\gamma^{mnp}r\right)=\left(r\gamma^{m}\gamma^{n}\gamma^{p}r\right)$
and $\left(\overline{\lambda}\gamma^{mnp}r\right)\left(r\gamma_{p}\right)^{\alpha}=\left(r\gamma^{mnp}r\right)\left(\overline{\lambda}\gamma_{p}\right)^{\alpha}$,
\begin{equation}
\left(\overline{\lambda}\gamma_{mnp}r\right)\left(\overline{\lambda}\gamma_{qrs}r\right)\left(r\gamma^{pqt}r\right)=\left(r\gamma_{mnp}r\right)\left(r\gamma_{qrs}r\right)\left(\overline{\lambda}\gamma^{p}\gamma^{t}\gamma^{q}\overline{\lambda}\right),
\end{equation}
which vanishes because $\left(\overline{\lambda}\gamma^{mnp}\overline{\lambda}\right)=0$.
\item terms with one $N$, one $\Pi$ and one partial derivative (Taylor
expansion of a quadratic pole), as
\begin{equation}
\frac{\left(\partial\overline{\lambda}\gamma_{mnp}r\right)\left(\overline{\lambda}\gamma_{qrs}r\right)\left(r\gamma^{pqr}r\right)N^{mn}\Pi^{s}}{\left(\overline{\lambda}\lambda\right)^{6}},
\end{equation}
which vanishes, since
\begin{eqnarray}
\left(\overline{\lambda}\gamma_{qrs}r\right)\left(r\gamma^{pqr}r\right) & = & \left(\overline{\lambda}\gamma_{qr}\gamma_{s}r\right)\left(r\gamma^{qr}\gamma^{p}r\right)\nonumber \\
 & = & 4\left(\overline{\lambda}\gamma^{m}r\right)\left(r\gamma_{s}\gamma_{m}\gamma^{p}r\right)\nonumber \\
 & - & 2\left(\overline{\lambda}\gamma_{s}r\right)\left(r\gamma^{p}r\right)-8\left(r\gamma_{s}r\right)\left(\overline{\lambda}\gamma^{p}r\right)\nonumber \\
 & = & 0.
\end{eqnarray}

\item terms with one $N$ and two $d$'s, like
\begin{equation}
\frac{\left(\overline{\lambda}\gamma_{mnp}r\right)\left(\overline{\lambda}\gamma^{mqr}r\right)N_{\phantom{m}r}^{n}\left(r\gamma^{p}d\right)\left(r\gamma_{q}d\right)}{\left(\overline{\lambda}\lambda\right)^{6}}.
\end{equation}
Since $\overline{\lambda}\gamma^{mnp}r$ is equal to $\overline{\lambda}\gamma^{m}\gamma^{n}\gamma^{p}r$,
this term is proportional to $\left(\overline{\lambda}\gamma^{m}\right)^{\alpha}\left(\overline{\lambda}\gamma_{m}\right)^{\beta}$,
and, according to equation \eqref{eq:gammaid7}, it vanishes.
\item terms with two $d$'s and one partial derivative, such as
\begin{equation}
\frac{\left(\overline{\lambda}\gamma_{mnp}r\right)\left(\partial\overline{\lambda}\gamma^{mnq}r\right)\left(r\gamma^{p}d\right)\left(r\gamma_{q}d\right)}{\left(\overline{\lambda}\lambda\right)^{6}}.
\end{equation}
Decomposing $\left(\partial\overline{\lambda}\gamma^{mnp}r\right)$
as $\left(\partial\overline{\lambda}\gamma^{mn}\gamma^{p}r\right)+\eta^{np}\left(\overline{\lambda}\gamma^{m}\partial r\right)-\eta^{mp}\left(\overline{\lambda}\gamma^{n}\partial r\right)$,
it is possible to rewrite the expression as follows,
\begin{eqnarray}
\left(\overline{\lambda}\gamma_{mnp}r\right)\left(\partial\overline{\lambda}\gamma^{mnq}r\right) & = & \left(\overline{\lambda}\gamma_{mn}\gamma_{p}r\right)\left(\partial\overline{\lambda}\gamma^{mn}\gamma^{q}r\right)+2\eta^{nq}\left(\overline{\lambda}\gamma_{mnp}r\right)\left(\overline{\lambda}\gamma^{m}\partial r\right)\nonumber \\
 & = & \left(\overline{\lambda}\gamma_{m}\partial\overline{\lambda}\right)\left(r\gamma_{p}\gamma_{m}\gamma^{q}r\right)-2\left(\overline{\lambda}\gamma_{p}r\right)\left(\partial\overline{\lambda}\gamma^{q}r\right)\nonumber \\
 & - & 8\left(\overline{\lambda}\gamma^{q}r\right)\left(\partial\overline{\lambda}\gamma_{p}r\right)+2\eta^{nq}\left(\overline{\lambda}\gamma_{m}\gamma_{np}r\right)\left(\overline{\lambda}\gamma^{m}\partial r\right)\nonumber \\
 & = & 0,
\end{eqnarray}
showing that this term also vanishes.
\item and terms with one $\Pi$ and two partial derivatives (Taylor expansion
of a cubic pole), like
\begin{equation}
\frac{\left(\partial\overline{\lambda}\gamma_{mnp}\partial r\right)\left(\overline{\lambda}\gamma^{mnq}r\right)\left(r\gamma_{\phantom{p}qr}^{p}r\right)\Pi^{r}}{\left(\overline{\lambda}\lambda\right)^{6}}.
\end{equation}
Decomposing $\left(\partial\overline{\lambda}\gamma^{mnp}\partial r\right)$
as $\left(\partial\overline{\lambda}\gamma^{mn}\gamma^{p}\partial r\right)-\eta^{np}\left(\partial\overline{\lambda}\gamma^{m}\partial r\right)+\eta^{mp}\left(\partial\overline{\lambda}\gamma^{n}\partial r\right)$,
the expression
\begin{equation}
\left(\partial\overline{\lambda}\gamma_{mnp}\partial r\right)\left(\overline{\lambda}\gamma^{mnq}r\right)\left(r\gamma_{\phantom{p}qr}^{p}r\right)
\end{equation}
can be split into two pieces. One of them is similar to the ones presented
before and also vanishes. The other one is proportional to
\begin{eqnarray}
\left(\overline{\lambda}\gamma_{m}\partial r\right)\left(\overline{\lambda}\gamma_{n}\partial r\right)\left(r\gamma^{mnp}r\right) & = & \left(r\gamma_{m}\partial r\right)\left(\overline{\lambda}\gamma_{n}\partial r\right)\left(\overline{\lambda}\gamma^{mnp}r\right)\nonumber \\
 & = & -\left(r\gamma_{m}\partial r\right)\left(\overline{\lambda}\gamma_{n}\partial r\right)\left(\overline{\lambda}\gamma^{n}\gamma^{mp}r\right),
\end{eqnarray}
and vanishes, since $\left(\overline{\lambda}\gamma^{m}\right)^{\alpha}\left(\overline{\lambda}\gamma_{m}\right)^{\beta}=0$.
\end{itemize}
\

For
\begin{equation}
\left(bb\right)_{5}\equiv\int dz\left\{ b_{3}\left(z\right)b_{2}\left(y\right)+b_{2}\left(z\right)b_{3}\left(y\right)\right\} ,
\end{equation}
all contributions to the simple pole will have $d_{\alpha}$:
\begin{itemize}
\item there are terms with two $N$'s, as
\begin{equation}
\frac{\left(\overline{\lambda}\gamma_{mnp}r\right)\left(r\gamma^{p}d\right)\left(\overline{\lambda}\gamma_{qrs}r\right)\left(r\gamma^{stu}r\right)N^{mq}\eta^{nr}N_{tu}}{\left(\overline{\lambda}\lambda\right)^{7}}.
\end{equation}
Note that
\begin{eqnarray}
\left(\overline{\lambda}\gamma_{mnp}r\right)\left(\overline{\lambda}\gamma_{qrs}r\right)\eta^{nr} & = & \left(\overline{\lambda}\gamma_{n}\gamma_{mp}r\right)\left(\overline{\lambda}\gamma_{r}\gamma_{qs}r\right)\eta^{nr}\nonumber \\
 & = & \left(\overline{\lambda}\gamma_{m}\right)^{\alpha}\left(\overline{\lambda}\gamma^{m}\right)^{\beta}\left(\ldots\right)_{\alpha\beta}\nonumber \\
 & = & 0,
\end{eqnarray}
gives a vanishing contribution.
\item terms with one $N$ and one partial derivative, as
\begin{equation}
\frac{\left(\overline{\lambda}\gamma_{mnp}\partial r\right)\left(r\gamma^{p}d\right)\left(\overline{\lambda}\gamma^{mnq}r\right)\left(r\gamma_{qrs}r\right)N^{rs}}{\left(\overline{\lambda}\lambda\right)^{7}}.
\end{equation}
It is easy to extract the pure spinor constraint out of this expression:
\begin{eqnarray}
\left(\overline{\lambda}\gamma_{mnp}\partial r\right)\left(\overline{\lambda}\gamma^{mnq}r\right) & = & \left(\overline{\lambda}\gamma_{mn}\gamma_{p}\partial r\right)\left(\overline{\lambda}\gamma^{mn}\gamma^{q}r\right)\nonumber \\
 & - & 2\left(\overline{\lambda}\gamma_{m}\partial r\right)\left(\overline{\lambda}\gamma^{m}\gamma^{nq}r\right)\eta_{np}\nonumber \\
 & = & 4\left(\overline{\lambda}\gamma^{m}\overline{\lambda}\right)\left(\partial r\gamma_{p}\gamma_{m}\gamma^{q}r\right)-10\left(\overline{\lambda}\gamma_{p}\partial r\right)\left(\overline{\lambda}\gamma^{q}r\right)\nonumber \\
 & - & 2\left(\overline{\lambda}\gamma_{m}\partial r\right)\left(\overline{\lambda}\gamma^{m_{0}}\gamma^{nq}r\right)\eta_{np}.\nonumber \\
 & = & 0.
\end{eqnarray}

\item and terms with two partial derivatives, coming from the cubic poles,
like
\begin{equation}
\frac{\left(\partial\overline{\lambda}\gamma_{mnp}\partial r\right)\left(r\gamma^{p}d\right)\left(\overline{\lambda}\gamma^{mnq}r\right)\left(r\gamma_{qrs}r\right)\left(\overline{\lambda}\gamma^{rs}\lambda\right)}{\left(\overline{\lambda}\lambda\right)^{8}}.
\end{equation}
Note that $\left(r\gamma_{qrs}r\right)\left(\overline{\lambda}\gamma^{rs}\lambda\right)$
has the same structure of \eqref{eq:cubicpolevanishing} and also
vanishes.
\end{itemize}
\

Finally, for the last term in the $b\left(z\right)b\left(y\right)$
OPE, where only the ghost fields appear,
\begin{equation}
\left(bb\right)_{6}\equiv\int dz\left\{ b_{3}\left(z\right)b_{3}\left(y\right)\right\} ,
\end{equation}

\begin{itemize}
\item there are terms with three $N$'s, like
\begin{equation}
\frac{\left(\overline{\lambda}\gamma_{mnp}r\right)\left(r\gamma^{pqr}r\right)\left(\overline{\lambda}\gamma^{mst}r\right)\left(r\gamma_{tuv}r\right)N_{qr}N_{\phantom{m_{0}}s}^{n}N^{uv}}{\left(\overline{\lambda}\lambda\right)^{8}}.
\end{equation}
Since $\overline{\lambda}\gamma^{mnp}r=\overline{\lambda}\gamma^{m}\gamma^{n}\gamma^{p}r$,
$\left(\overline{\lambda}\gamma_{mnp}r\right)\left(\overline{\lambda}\gamma^{mqr}r\right)$
vanishes, as shown above.
\item terms with two $N$'s and one partial derivative, like
\begin{equation}
\frac{\partial\left(\overline{\lambda}\gamma_{mnp}r\right)\left(r\gamma^{pqr}r\right)\left(\overline{\lambda}\gamma^{mns}r\right)\left(r\gamma_{stu}r\right)N_{qr}N^{tu}}{\left(\overline{\lambda}\lambda\right)^{8}},
\end{equation}
which has the same structure presented before, being proportional
to the pure spinor constraints.
\item terms with one $N$ and two partial derivatives, coming from triple
poles, such as
\begin{equation}
\frac{\partial^{2}\left(\overline{\lambda}\gamma_{mnp}r\right)\left(r\gamma^{pqr}r\right)\left(\overline{\lambda}\gamma^{mns}r\right)\left(r\gamma_{qst}r\right)N_{r}^{\phantom{r}t}}{\left(\overline{\lambda}\lambda\right)^{8}},
\end{equation}
which are similar to the above ones and vanish.
\item and terms with three partial derivatives, like
\begin{equation}
\frac{\left(\partial\overline{\lambda}\gamma_{mnp}\partial r\right)\left(r\gamma^{pqr}\partial r\right)\left(\overline{\lambda}\gamma^{mns}r\right)\left(r\gamma_{qrs}r\right)}{\left(\overline{\lambda}\lambda\right)^{8}},
\end{equation}
that can be rewritten as
\begin{equation}
\frac{\left(\partial\overline{\lambda}\gamma_{m}\partial r\right)\left(r\gamma^{q}\partial r\right)\left(\overline{\lambda}\gamma^{mnp}r\right)\left(r\gamma_{npq}r\right)}{\left(\overline{\lambda}\lambda\right)^{8}}
\end{equation}
and vanish, since
\begin{eqnarray}
\left(\overline{\lambda}\gamma^{mnp}r\right)\left(r\gamma_{npq}r\right) & = & \left(\overline{\lambda}\gamma^{np}\gamma^{m}r\right)\left(r\gamma_{np}\gamma_{q}r\right)\nonumber \\
 & = & \left(\overline{\lambda}\gamma^{n}r\right)\left(r\gamma^{m}\gamma_{n}\gamma_{q}r\right)\nonumber \\
 & - & 2\left(\overline{\lambda}\gamma^{m}r\right)\left(r\gamma_{q}r\right)-8\left(r\gamma^{m}r\right)\left(\overline{\lambda}\gamma_{q}r\right)\nonumber \\
 & = & 0.
\end{eqnarray}

\end{itemize}
Summarizing, in the OPE computation several terms vanish identically
due to the pure spinor constraints (in particular, $\left(bb\right)_{4}$,
$\left(bb\right)_{5}$ and $\left(bb\right)_{6}$ do not present nontrivial
contributions). The remaining terms are excluded through the BRST
argument, since they were shown to be not BRST closed. Therefore,
\begin{equation}
\left(bb\right)_{1}=\left(bb\right)_{2}=\left(bb\right)_{3}=\left(bb\right)_{4}=\left(bb\right)_{5}=\left(bb\right)_{6}=0,
\end{equation}
and the pure spinor $b$ ghost is, indeed, nilpotent:
\begin{equation}
b\left(z\right)b\left(y\right)\sim\textrm{regular}.\label{eq:OPEbb}
\end{equation}

\section{Conclusion}

\

In this work, some properties of the $b$ ghost in the non-minimal
pure spinor formalism were reviewed and confirmed. The main object
of study was the nilpotency of the non-minimal $b$ ghost.

From general arguments, the $b\left(z\right)b\left(y\right)$ OPE
is reduced to
\[
b\left(z\right)b\left(y\right)\sim\frac{O_{1}}{\left(z-y\right)^{3}}+\frac{1}{2}\frac{\partial O_{1}}{\left(z-y\right)^{2}}+\frac{O_{3}}{\left(z-y\right)},
\]
where $O_{1}$ and $O_{3}$ are BRST closed.

As was already known from \cite{Chandia:2010ix}, the different terms
in the cubic pole, $O_{1}$, are all proportional to the pure spinor
constraints
\[
\overline{\lambda}\gamma^{m}r=\overline{\lambda}\gamma^{m}\overline{\lambda}=0.
\]
However, the demonstration that the simple pole ($O_{3}$) vanishes,
was incomplete, due to a wrong assumption on the absence of $r_{\alpha}$
derivatives.

A counter-example to that assumption is very simple,
\[
\frac{\left(\overline{\lambda}\gamma_{m}\partial r\right)\left(r\gamma^{m}\partial r\right)\left(\overline{\lambda}\partial\theta\right)}{\left(\overline{\lambda}\lambda\right)^{5}}.
\]
Note that the fundamental ingredient here is $\left(r\gamma^{m}\partial r\right)$,
an object that does not allow, in general, the removal of the partial
derivatives acting on $r_{\alpha}$.

Knowing this flaw, the proof that $O_{3}=0$ was carried out in a
straightforward manner. First, a careful analysis was made, obtaining
all terms that could be generated in the OPE computation. For some
of them, the cancellation is very simple to obtain and no BRST argument
is needed. However, for most of the terms (those that appear in ordering
rearrangements, for example), a direct check is very hard to perform.
However, they were shown to be not BRST closed. Since the possible
poles appearing in $b\left(z\right)b\left(y\right)$ must commute
with the BRST charge, $O_{3}$ vanishes, and this constitutes a rigorous
proof of the $b$ ghost nilpotency in the non-minimal pure spinor
formalism, confirming the interpretation of the theory as a $\hat{c}=3$
$\mathcal{N}=2$ topological string.

\

\textbf{Acknowledgements:} I would like to thank Nathan Berkovits,
Ido Adam and Ilya Bakhmatov for useful discussions. Also, for reading
the manuscript, Thales Agricola and Chrysostomos Kalousios. This work
was supported by FAPESP grant 2009/17516-4.

\appendix

\section{Conventions and useful formulas\label{sec:Conventions}}

\subsection*{Conventions}

\

Indices:
\[
\begin{cases}
m,n,\ldots=0,\ldots,9 & \textrm{space-time vector indices,}\\
\alpha,\beta,\ldots=1,\ldots,16 & \textrm{space-time spinor indices,}
\end{cases}
\]

The indices antisymmetrization is represented by the square brackets,
meaning
\begin{equation}
\left[I_{1}\ldots I_{n}\right]=\frac{1}{n!}\left(I_{1}\ldots I_{n}+\textrm{all antisymmetric permutations}\right).
\end{equation}
For example,
\begin{equation}
\gamma^{[m}\gamma^{n]}=\frac{1}{2}\left(\gamma^{m}\gamma^{n}-\gamma^{n}\gamma^{m}\right)=\gamma^{mn},
\end{equation}
or,
\begin{equation}
\lambda^{[\alpha}H^{\beta\gamma]}=\frac{1}{3!}\left(\lambda^{\alpha}H^{\beta\gamma}-\lambda^{\alpha}H^{\gamma\beta}+\lambda^{\beta}H^{\gamma\alpha}-\lambda^{\beta}H^{\alpha\gamma}+\lambda^{\gamma}H^{\alpha\beta}-\lambda^{\gamma}H^{\beta\alpha}\right).
\end{equation}

Concerning OPE's, the right-hand sides of the equations are always
evaluated at the coordinate of the second entry, that is,
\begin{equation}
A\left(z\right)B\left(y\right)\sim\frac{C}{\left(z-y\right)^{2}}+\frac{D}{\left(z-y\right)}
\end{equation}
means $C=C\left(y\right)$ and $D=D\left(y\right)$.

\subsection*{Gamma matrices}

\

The gamma matrices $\gamma_{\alpha\beta}^{m}$ and $\gamma_{m}^{\alpha\beta}$
satisfy
\begin{equation}
\left\{ \gamma^{m},\gamma^{n}\right\} _{\phantom{\alpha}\beta}^{\alpha}=\left(\gamma^{m}\right)^{\alpha\sigma}\gamma_{\sigma\beta}^{n}+\left(\gamma^{n}\right)^{\alpha\sigma}\gamma_{\sigma\beta}^{m}=2\eta^{mn}\delta_{\beta}^{\alpha}.\label{eq:diracalgebra}
\end{equation}

The Fierz decompositions of bispinors are given by\begin{subequations}\label{eq:Fierz}
\begin{eqnarray}
\chi^{\alpha}\psi^{\beta} & = & \frac{1}{16}\gamma_{m}^{\alpha\beta}\left(\chi\gamma^{m}\psi\right)+\frac{1}{3!16}\gamma_{mnp}^{\alpha\beta}\left(\chi\gamma^{mnp}\psi\right)+\frac{1}{5!16}\left(\frac{1}{2}\right)\gamma_{mnpqr}^{\alpha\beta}\left(\chi\gamma^{mnpqr}\psi\right),\\
\chi_{\alpha}\psi^{\beta} & = & \frac{1}{16}\delta_{\alpha}^{\beta}\left(\chi\psi\right)-\frac{1}{2!16}\left(\gamma_{mn}\right)_{\phantom{\beta}\alpha}^{\beta}\left(\chi\gamma^{mn}\psi\right)+\frac{1}{4!16}\left(\gamma_{mnpq}\right)_{\phantom{\beta}\alpha}^{\beta}\left(\chi\gamma^{mnpq}\psi\right),
\end{eqnarray}
\end{subequations}where
\begin{equation}
\begin{array}{ccc}
\gamma_{m}^{\alpha\beta}=\gamma_{m}^{\beta\alpha}, & \gamma_{mnp}^{\alpha\beta}=-\gamma_{mnp}^{\beta\alpha}, & \gamma_{mnpqr}^{\alpha\beta}=\gamma_{mnpqr}^{\beta\alpha}.\end{array}
\end{equation}

The main gamma matrix identity that is being used in this work is
\begin{equation}
\left(\gamma^{mn}\right)_{\phantom{\alpha}\beta}^{\alpha}\left(\gamma_{mn}\right)_{\phantom{\gamma}\lambda}^{\gamma}=4\gamma_{\beta\lambda}^{m}\gamma_{m}^{\alpha\gamma}-2\delta_{\beta}^{\alpha}\delta_{\lambda}^{\gamma}-8\delta_{\lambda}^{\alpha}\delta_{\beta}^{\gamma},\label{eq:gammaid1}
\end{equation}
which can be deduced from \eqref{eq:Fierz}. The other relevant one
is given by
\begin{equation}
\eta_{mn}\left(\gamma_{\alpha\beta}^{m}\gamma_{\gamma\lambda}^{n}+\gamma_{\alpha\gamma}^{m}\gamma_{\beta\lambda}^{n}+\gamma_{\alpha\lambda}^{m}\gamma_{\gamma\beta}^{n}\right)=0.\label{eq:gammaid2}
\end{equation}

There are several other identities that can be derived from \eqref{eq:gammaid1}:
\begin{eqnarray}
 & \left(\gamma^{mn}\right)_{\phantom{\alpha}\beta}^{\alpha}\gamma_{mnp}^{\gamma\lambda}=2\left(\gamma^{m}\right)^{\alpha\gamma}\left(\gamma_{pm}\right)_{\phantom{\alpha}\beta}^{\lambda}+6\gamma_{p}^{\alpha\gamma}\delta_{\beta}^{\lambda}-\left(\gamma\leftrightarrow\lambda\right),\label{eq:gammaid3}\\
 & \left(\gamma_{mn}\right)_{\phantom{\alpha}\beta}^{\alpha}\gamma_{\gamma\lambda}^{mnp}=-2\left(\gamma_{m}\right)_{\beta\lambda}\left(\gamma^{pm}\right)_{\phantom{\alpha}\gamma}^{\alpha}+6\gamma_{\beta\lambda}^{p}\delta_{\gamma}^{\alpha}-\left(\gamma\leftrightarrow\lambda\right),\label{eq:gammaid4}\\
 & \gamma_{mnp}^{\alpha\beta}\left(\gamma^{mnp}\right)^{\gamma\lambda}=12\left[\gamma_{m}^{\alpha\lambda}\left(\gamma^{m}\right)^{\beta\gamma}-\gamma_{m}^{\alpha\gamma}\left(\gamma^{m}\right)^{\beta\lambda}\right],\label{eq:gammaid5}\\
 & \gamma_{mnp}^{\alpha\beta}\gamma_{\gamma\lambda}^{mnp}=48\left(\delta_{\gamma}^{\alpha}\delta_{\lambda}^{\beta}-\delta_{\lambda}^{\alpha}\delta_{\gamma}^{\beta}\right).\label{eq:gammaid6}
\end{eqnarray}
All of them are very helpful in extracting the pure spinor constraints
out of product of bispinors containing space-time vector indices contracted.
For example:
\begin{eqnarray}
\left(\overline{\lambda}\gamma_{mnp}r\right)\left(\overline{\lambda}\gamma^{mn}\lambda\right) & = & 2\left(\overline{\lambda}\gamma^{m}\overline{\lambda}\right)\left(r\gamma_{pm}\lambda\right)+6\left(r\lambda\right)\left(\overline{\lambda}\gamma_{p}\overline{\lambda}\right)\nonumber \\
 &  & -2\left(\overline{\lambda}\gamma^{m}r\right)\left(\overline{\lambda}\gamma_{pm}\lambda\right)-6\left(\overline{\lambda}\lambda\right)\left(\overline{\lambda}\gamma_{p}r\right)\nonumber \\
 & = & 0.
\end{eqnarray}

The last identity that is often used in the calculations is
\begin{equation}
\gamma^{m}\gamma^{n_{1}\ldots n_{k}}\gamma_{m}=\left(-1\right)^{k}\left(10-2k\right)\gamma^{n_{1}\ldots n_{k}},\label{eq:gammaid7}
\end{equation}
which is particularly useful since it implies that $\left(\gamma^{m}\lambda\right)_{\alpha}\left(\gamma_{m}\lambda\right)_{\beta}=0$
for $\lambda$ being a pure spinor.

\subsection*{Ordering considerations}

\

This part intended to present some aspects of the ordering prescription
that is being used in this work.

Classical relations between currents are now corrected with ordering
contributions. For example,
\begin{equation}
N_{\textrm{cl}}^{mn}\left(\gamma_{n}\lambda\right)_{\alpha}=\frac{1}{2}J_{\textrm{cl}}\left(\gamma^{m}\lambda\right)_{\alpha}
\end{equation}
is valid for any pure spinor $\lambda$. Its quantum version is given
by
\begin{equation}
\left(N^{mn},\lambda^{\beta}\right)\gamma_{\alpha\beta}^{p}\eta_{np}-\frac{1}{2}\left(J_{\lambda},\lambda^{\beta}\right)\gamma_{\alpha\beta}^{m}=2\left(\gamma^{m}\partial\lambda\right)_{\alpha},
\end{equation}
showing that some of the $45$ Lorentz generators can be written in
terms of the others (in fact, only $10$ are independent components).

Another important example is the equation
\begin{equation}
4\lambda^{\alpha}T_{\textrm{cl}}+J_{\textrm{cl}}\partial\lambda^{\alpha}+N_{\textrm{cl}}^{mn}\left(\gamma_{mn}\partial\lambda\right)^{\alpha}=0,
\end{equation}
which establishes a connection between the energy-momentum tensor
and the other currents. Implementing the ordering leads to

\begin{equation}
\left(\lambda^{\alpha},T\right)+4\partial^{2}\lambda^{\alpha}=-\frac{1}{4}\left(J_{\lambda},\partial\lambda^{\alpha}\right)-\frac{1}{4}\left(N_{mn},\left(\gamma^{mn}\partial\lambda\right)^{\alpha}\right).
\end{equation}
This relation appears in the construction of the quantum $b$ ghost,
as well as

\begin{equation}
\left(\frac{1}{4}\right)\gamma_{mnp}^{\beta\alpha}\left(N^{mn},\lambda\gamma^{p}\partial\theta\right)=8\partial\lambda^{[\alpha}\partial\theta^{\beta]}+\left(\lambda^{[\alpha},N_{mn}\left(\gamma^{mn}\partial\theta\right)^{\beta]}\right)+\left(\lambda^{[\alpha},J_{\lambda}\partial\theta^{\beta]}\right),
\end{equation}
which is the ordered version of
\begin{equation}
\gamma_{mnp}^{\alpha\beta}N_{\textrm{cl}}^{mn}\left(\lambda\gamma^{p}\partial\theta\right)+4\lambda^{[\alpha}N_{\textrm{cl}}^{mn}\left(\gamma_{mn}\partial\theta\right)^{\beta]}+4\lambda^{[\alpha}J_{\textrm{cl}}\partial\theta^{\beta]}=0.
\end{equation}

A further application is the Sugawara construction of the energy-momentum
tensor for the minimal ghost sector,
\begin{equation}
T_{\lambda}=-\frac{1}{20}\left(N^{mn},N_{mn}\right)-\frac{1}{8}\left(J_{\lambda},J_{\lambda}\right)+\partial J_{\lambda},\label{eq:sugawaraTmin}
\end{equation}
which correctly reproduces the related OPE's.

OPE computations are more systematic%
\footnote{See chapter $6$ of \cite{DiFrancesco:1997nk}, where the normal ordering
is presented in details.%
} within the prescription \eqref{eq:ordering}. As an example, it will
be shown here that the $b$ ghost for the non-minimal formalism is
a primary field.

Concerning $b_{-1}$, the ordering does not matter and it is straightforward
to see that
\begin{equation}
T\left(z\right)b_{-1}\left(y\right)\sim2\frac{b_{-1}}{\left(z-y\right)^{2}}+\frac{\partial b_{-1}}{\left(z-y\right)}.
\end{equation}

For $b_{0}$, however, there are some subtleties. Analyzing $G^{\alpha}$
first,
\begin{equation}
T\left(z\right)G^{\alpha}\left(y\right)\sim2\frac{G^{\alpha}}{\left(z-y\right)^{2}}+\frac{\partial G^{\alpha}}{\left(z-y\right)}+\frac{\partial\theta^{\alpha}}{\left(z-y\right)^{3}}.\label{eq:OPETGalpha}
\end{equation}
Note that the cubic pole receives contributions from $J_{\lambda}$
(the ghost current anomaly), $\partial^{2}\theta^{\alpha}$ and $\left(\Pi^{m},\gamma_{m}^{\alpha\beta}d_{\beta}\right)$:
\begin{eqnarray}
T\left(z\right)J_{\lambda}\left(y\right) & \sim & \frac{8}{\left(z-y\right)^{3}}+\frac{J_{\lambda}}{\left(z-y\right)^{2}}+\frac{\partial J_{\lambda}}{\left(z-y\right)},\\
T\left(z\right)\partial^{2}\theta^{\alpha}\left(y\right) & \sim & 2\frac{\partial\theta^{\alpha}}{\left(z-y\right)^{3}}+2\frac{\partial^{2}\theta^{\alpha}}{\left(z-y\right)^{2}}+\frac{\partial^{3}\theta^{\alpha}}{\left(z-y\right)},\\
T\left(z\right)\left(\Pi^{m},\gamma_{m}^{\alpha\beta}d_{\beta}\right)\left(y\right) & \sim & \left(\frac{\Pi^{m}}{\left(z-y\right)^{2}}+\frac{\partial\Pi^{m}}{\left(z-y\right)},\gamma_{m}^{\alpha\beta}d_{\beta}\right)\nonumber \\
 & + & \left(\Pi^{m},\frac{\gamma_{m}^{\alpha\beta}d_{\beta}}{\left(z-y\right)^{2}}+\frac{\gamma_{m}^{\alpha\beta}\partial d_{\beta}}{\left(z-y\right)}\right).
\end{eqnarray}
According to the ordering prescription, the first term in the last
OPE can be rewritten as
\begin{multline}
\frac{1}{2\pi i}\oint\frac{dw}{\left(w-y\right)}\left\{ \frac{1}{\left(z-w\right)^{2}}\Pi^{m}\left(w\right)+\frac{1}{\left(z-w\right)}\partial\Pi^{m}\left(w\right)\right\} \gamma_{m}^{\alpha\beta}d_{\beta}\left(y\right)=\\
-10\frac{\partial\theta^{\alpha}}{\left(z-y\right)^{3}}+\frac{\left(\Pi^{m},\gamma_{m}^{\alpha\beta}d_{\beta}\right)}{\left(z-y\right)^{2}}+\frac{\left(\partial\Pi^{m},\gamma_{m}^{\alpha\beta}d_{\beta}\right)}{\left(z-y\right)}.
\end{multline}
Therefore,
\begin{equation}
T\left(z\right)\left(\Pi^{m},\gamma_{m}^{\alpha\beta}d_{\beta}\right)\left(y\right)\sim-10\frac{\partial\theta^{\alpha}}{\left(z-y\right)^{3}}+2\frac{\left(\Pi^{m},\gamma_{m}^{\alpha\beta}d_{\beta}\right)}{\left(z-y\right)^{2}}+\frac{\partial\left(\Pi^{m},\gamma_{m}^{\alpha\beta}d_{\beta}\right)}{\left(z-y\right)}.
\end{equation}
Adding up all the contributions, equation \eqref{eq:OPETGalpha} is
reproduced. For the whole $b_{0}$,

\begin{multline}
T\left(z\right)b_{0}\left(y\right)\sim\left(\frac{1}{\left(z-y\right)}\partial\left(\frac{\overline{\lambda}_{\alpha}}{\overline{\lambda}\lambda}\right),G^{\alpha}\right)\\
+\left(\frac{\overline{\lambda}_{\alpha}}{\overline{\lambda}\lambda},2\frac{G^{\alpha}}{\left(z-y\right)^{2}}+\frac{\partial G^{\alpha}}{\left(z-y\right)}+\frac{\partial\theta^{\alpha}}{\left(z-y\right)^{3}}\right)+2\frac{O}{\left(z-y\right)^{2}}+\frac{\partial O}{\left(z-y\right)}.\label{eq:OPETG}
\end{multline}
Again, the first term on the right-hand side can be rewritten as
\begin{multline}
\frac{1}{2\pi i}\oint\frac{dw}{\left(w-y\right)}\frac{1}{\left(z-w\right)}\partial_{w}\left(\frac{\overline{\lambda}_{\alpha}}{\overline{\lambda}\lambda}\right)\left(w\right)G^{\alpha}\left(y\right)=\\
=\frac{1}{\left(z-y\right)}\left(\partial\left(\frac{\overline{\lambda}_{\alpha}}{\overline{\lambda}\lambda}\right),G^{\alpha}\right)-\frac{1}{\left(z-y\right)^{3}}\left(\frac{\overline{\lambda}_{\alpha}\partial\theta^{\alpha}}{\overline{\lambda}\lambda}\right).
\end{multline}
Replacing this equation in \eqref{eq:OPETG}, the cubic pole disappears,
yielding a primary field.

For $b_{1}$, $b_{2}$ and $b_{3}$, there are no contributions like
the one in $b_{0}$ (they are all proportional to the pure spinor
constraints), therefore the $b$ ghost, given by equation \eqref{eq:quantumb},
is a primary field:
\begin{equation}
T\left(z\right)b\left(y\right)\sim2\frac{b}{\left(z-y\right)^{2}}+\frac{\partial b}{\left(z-y\right)}.
\end{equation}

\

\end{document}